\documentclass[fleqn,10pt]{wlscirep}
\usepackage[utf8]{inputenc}
\usepackage[T1]{fontenc}

\usepackage{algorithm}
\usepackage[noend]{algpseudocode}

\newif\ifhyper
\hypertrue
\ifhyper
\hypersetup{
   citecolor = {red},
   colorlinks = {true}, 
   linkcolor = {blue},
   urlcolor = {blue} 
}
\fi

\def\be{\begin{equation}}
\def\ee{\end{equation}}
\def\bea{\begin{eqnarray}}
\def\eea{\end{eqnarray}}

\title{Variational Tensor Neural Networks for Deep Learning}

\author[1,2,3]{Saeed S. Jahromi}
\author[2,3,4,*]{Rom\'an Or\'us}

\affil[1]{Department of Physics, Institute for Advanced Studies in Basic Sciences (IASBS), Zanjan 45137-66731, Iran}
\affil[2]{Donostia International Physics Center, Paseo Manuel de Lardizabal 4, E-20018 San Sebasti\'an, Spain}
\affil[3]{Multiverse Computing, Paseo de Miram\'on 170, E-20014 San Sebasti\'an, Spain}
\affil[4]{Ikerbasque Foundation for Science, Maria Diaz de Haro 3, E-48013 Bilbao, Spain}

\affil[*]{roman.orus@dipc.org}

\begin{abstract}
Deep neural networks (NNs) encounter scalability limitations when confronted with a vast array of neurons, thereby constraining their achievable network depth. To address this challenge, we propose an integration of tensor networks (TN) into NN frameworks, combined with a variational DMRG-inspired training technique. This in turn, results in a scalable tensor neural network (TNN) architecture capable of efficient training over a large parameter space. Our variational algorithm utilizes a  local gradient-descent technique, enabling manual or automatic computation of tensor gradients, facilitating design of hybrid TNN models with combined dense and tensor layers. Our training algorithm further provides insight on the entanglement structure of the tensorized trainable weights and correlation among the model parameters. We validate the accuracy and efficiency of our method by designing TNN models and providing benchmark results for linear and non-linear regressions, data classification and image recognition on MNIST handwritten digits.
\end{abstract}

\begin{document}

\flushbottom
\maketitle
%
%

\section*{Introduction}

Machine learning algorithms based on deep learning have demonstrated remarkable success in a diverse range of tasks, including identification \cite{Zhou2017,Sonoda2019,Chang2018}, classification \cite{Li2019,Affonso2017,Krizhevsky2017}, regression \cite{Lathuiliere2020, Ramsundar2018}, clustering \cite{Tian2017, Min2018, Aljalbout2018}, and numerous other artificial intelligence applications. Deep learning algorithms, particularly those employing neural networks (NNs), have garnered significant attention due to their enhanced capabilities in feature learning \cite{Yu2013,Jing2021} and decision-making \cite{Dehaene2000,Hill1994}, both in supervised and unsupervised learning scenarios. In addition, NNs have also found practical applications in condensed matter and statistical physics \cite{Bedolla2020, Karniadakis2021, Mehta2019}. As an example, it has been shown that NNs have expressive power for representing complex many-body wave functions \cite{Sharir2021,Pescia2022, Gutierrez2022,Pescia2022a, Wu2021}. Moreover, ideas for detecting phase transitions in quantum many-body systems with fully connected NNs and convolutional neural networks (CNN) have also been successfully put forward \cite{Carrasquilla2017, Greplova2020,Tanaka2017,Zhang2017}. 

Recently, the connection between tensor networks (TN), which are efficient ansatz for representing quantum many-body wave functions \cite{Orus2014,Verstraete2008}, and neural networks, has also been established. It has been shown that the trainable weights of NNs are closely related to many-body wave functions, henceforth can be replaced by TNs and trained using variational optimization techniques \cite{Stoudenmire2016,Patel2022,strashko_generalization_2022,barratt_improvements_2022}. Efficient TN algorithms for classification \cite{Stoudenmire2016}, anomaly detection \cite{Wang2020}, and clustering \cite{Stoudenmire2018} have been proposed. On top of their expressive power, TNs also provide efficient schemes for data compression based on tensor factorization \cite{Panagakis2021,Bahri2019,Kossaifi2017,Kossaifi2020}. As an example, the number of parameters in NN models can substantially be compressed by keeping only the most relevant degrees of freedom by discarding those that involve lower correlations. Tensor neural networks (TNN) \cite{Patel2022} and tensor convolutional neural networks (TCNN) \cite{MartinRamiro2022} are examples of NNs where the weight tensors of the hidden layers are replaced by tensor network structures using, e.g., the singular value decomposition (SVD). Recent studies actually confirm that, for such a reduced parameter space, TNNs have better performance and accuracy than standard NNs \cite{Patel2022, Panagakis2021}.

In most of the current TNN developments, the tensorization takes place only at the level of hidden layers (trainable weights) \cite{Patel2022, Kossaifi2017}. However, training of the model is generically performed by optimizing the contracted trainable weights of each layers based on standard optimization techniques such as gradient descent and automatic differentiation. Despite being efficient and accurate, these optimization schemes only target the global minimum of a loss function while being blind to the correlations and entanglement structure between the model parameters, in addition to being hard to scale. The behavior of a loss function monitors the training convergence in these approaches, and distinguishing local minima from actual global minima is in principle very difficult. Moreover, it is also difficult to infer meaningful information from such updated weight tensors.

In this paper we resolve these issues by fully integrating NNs with TNs. More specifically, we introduce an efficient TN layer in NN structures such that the trainable weights are represented by matrix product operators (MPO) \cite{Patel2022}. The TN layer can replace the fully connected (\texttt{Dense}) hidden layers of a NN. The resulting TNN is scalable and can have any desired number of TN layers to form a truly deep NN. We further introduce an entanglement-aware variational DMRG-like training algorithm for the resulting TNN. In contrast to previously-developed TN machine learning algorithms, which are single-layer and goal-specific \cite{Stoudenmire2016, Wang2020}, our TNN is an instance of a deep neural network that can be used for different data analysis tasks such as regression and classification. The DMRG-like training algorithm provides direct access to the entanglement spectrum of the MPO trainable weights, in turn giving rise to a clear insight on the correlations in the parameters of the machine learning model. Moreover, the entanglement structure of the MPOs and their expressive power as a quantum neural state can also be assessed by typical quantum information quantities such as, e.g., the entanglement entropy of a bipartition.

In order to demonstrate the efficiency and accuracy of our technique, we use our TN layer combined with different loss and activation functions to design different deep learning models. We use such algorithms for, e.g., linear and non-linear regression, classification as well as image recognition. Next, we use our DMRG-like algorithm \cite{White1992,White1993} to train the models. Our findings show that the TNNs have potential to be used as an alternative NN algorithm for different data analysis tasks with a general purpose. We also show how the non-linearity in the correlations amongst input data reflects itself in the entanglement spectrum and entanglement entropy of the MPO trainable weights. Our findings suggest that the DMRG-like training algorithm is an efficient numerical tool for machine learning that can potentially be used for detecting the TN representation of quantum neural states. 

The paper is organized as follows: In Sec.~\ref{sec:TNNs} we provide a brief overview of tensor networks and how a standard neural network composed of fully connected \texttt{Dense} layers can be tensorized to obtain a TNN. Next in Sec.~\ref{sec:dmrg} we introduce our DMRG-like training algorithm, together with the details of local tensor optimization updates based on gradient descent. In Sec.~\ref{sec:regressor} we build TNN models for regression and use them for fitting  linear- and non-linear random data. More examples of applications of TNNs for classification of labeled data are provided in Sec.~\ref{sec:classifier}. Additionally, benchmarks for image recognition of MNIST handwritten digits are presented in Sec.~\ref{sec:mnist}. Finally, in Sec.~\ref{sec:conclude} we wrap up with our conclusions and discussions about further possible work.

\section*{Tensor Neural Networks} 
\label{sec:TNNs} 
\begin{figure}
	\centerline{\includegraphics[width=12cm]{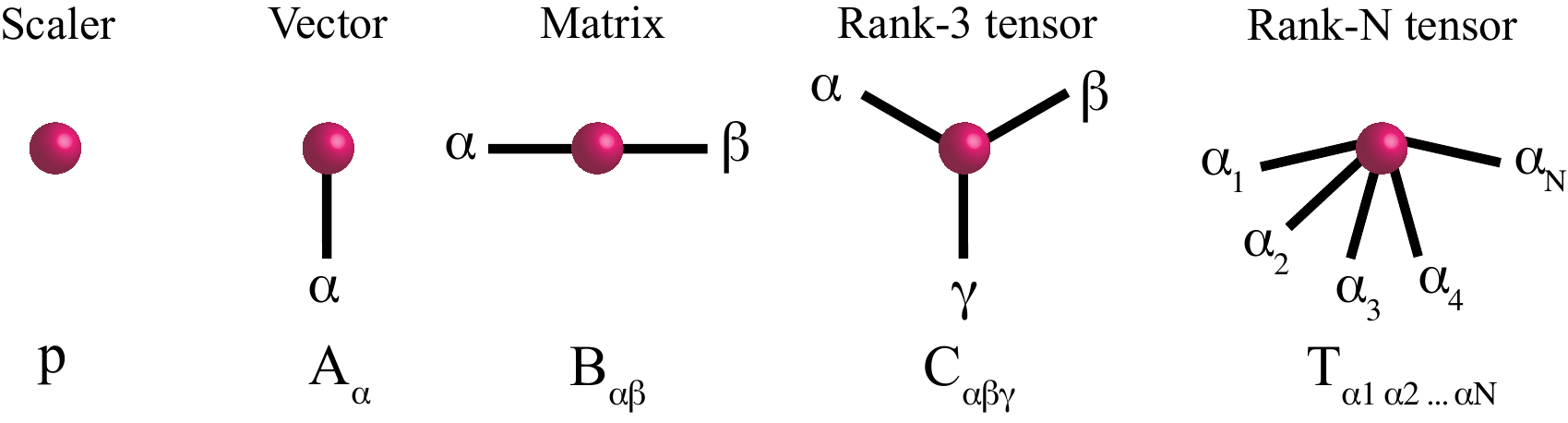}}
	\caption{[Color online] Diagrammatic representation of tensors. In this notation, a rank-$n$ tensor is represented by a solid object (here illustrated with spheres) with $n$ connected links (legs) so that a scalar, a vector, and a matrix are objects with zero, one, and two connected links, respectively. The number of links distinguishes the number of dimensions of the tensors. For example, a vector (matrix) is a one (two) dimensional array represented by a sphere with one (two) connected link(s).}
	\label{Fig:tn_diagrams}
\end{figure}

Tensor networks \cite{Orus2008,Verstraete2008} where originally developed in physics with the aim of providing efficient representations for quantum many-body wave functions. They are also the basis of well-established numerical techniques such as density matrix renormalization group (DMRG) \cite{White1992,White1993} and time-evolving block decimation (TEBD) \cite{Vidal2003,Vidal2004}. However, due to its high potential for efficient data representation and compression, novel applications of TNs are emerging in different branches of data science such as machine learning and optimization. In this section we show how TNs can enhance deep neural networks by providing an efficient representation of the trainable weights of classical neural networks. To this end, we first review some basic concepts on TNs in order to establish basic notation and concepts widely used in the physics' context. 

\subsection{Tensor Network Basics}
\label{sec:TN-basics} 
A tensor is a multi-dimensional array of complex numbers represented by $T_{\alpha\beta\gamma\ldots}$ in which the subscripts denote the tensor dimensions. The number of tensor dimensions further corresponds to the rank of the tensor. Tensors and tensor operations can alternatively be described by using tensor network diagrams \cite{Orus2014} as illustrated in Fig.~\ref{Fig:tn_diagrams}. Within this graphical notation, a rank-$n$ tensor is represented by a solid object (here illustrated by spheres) with $n$ connected links (legs) so that a scalar, a vector, and a matrix are objects with zero, one, and two connected links, respectively. This diagrammatic notation is then generalized to rank-$n$ tensors to object with $n$ legs, each corresponding to a tensor index. 

\begin{figure}
\centerline{\includegraphics[width=12cm]{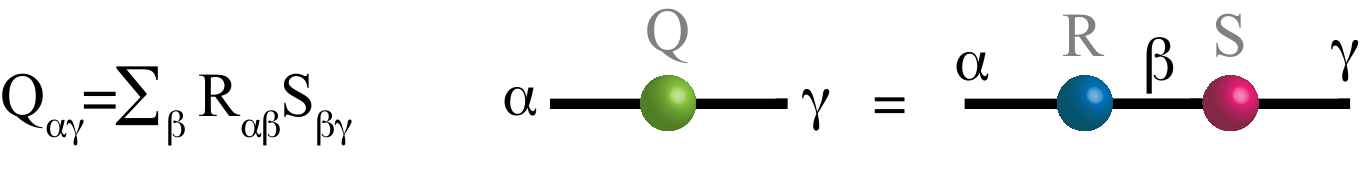}}
\caption{[Color online] Contraction of tensors, equivalent to tensor trace over their shared indices. This is represented graphically by connecting the shared links. Here, $R$ and $S$ tensors are connected along the shared leg $\beta$. This contraction operation is equivalent to matrix multiplication.}
\label{Fig:contraction}
\end{figure}

TN diagrams not only represent the tensors but also represent tensor contractions, which is the generalization of matrix multiplication to rank-$n$ tensors. For example, contraction of two rank-$2$ tensors, i.e., matrices $R_{\alpha\beta}$ and $S_{\beta\gamma}$ can be represented diagrammatically by connecting the two tensors along their shared link $\beta$, as shown in Fig.~\ref{Fig:contraction}. This operation can further be represented mathematically as
\be
Q_{\alpha\gamma} = {\rm tTR}(R_{\alpha\beta} S_{\beta\gamma}) = \sum_\beta R_{\alpha\beta} S_{\beta\gamma},
\ee
where the ${\rm tTR}$ is the tensor trace over shared indices (tensor legs). 

\begin{figure}
	\centerline{\includegraphics[width=7cm]{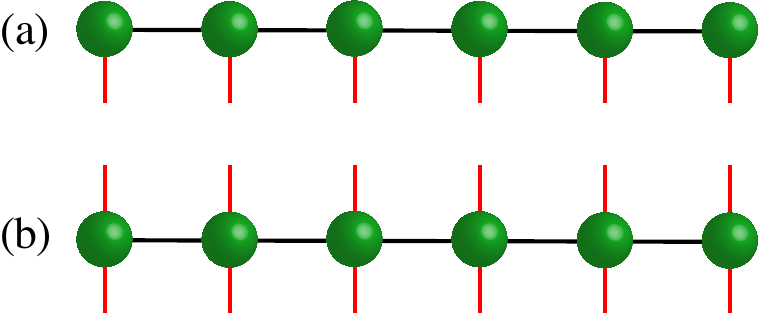}}
	\caption{[Color online] Examples of well-known one dimensional TNs. (a) Matrix product states (MPS) and (b) matrix product operators (MPO). The latter will be used for representing the trainable weights of \texttt{TNLayer}s.}
	\label{Fig:mps_mpo}
\end{figure}

\begin{figure}
	\centerline{\includegraphics[width=12cm]{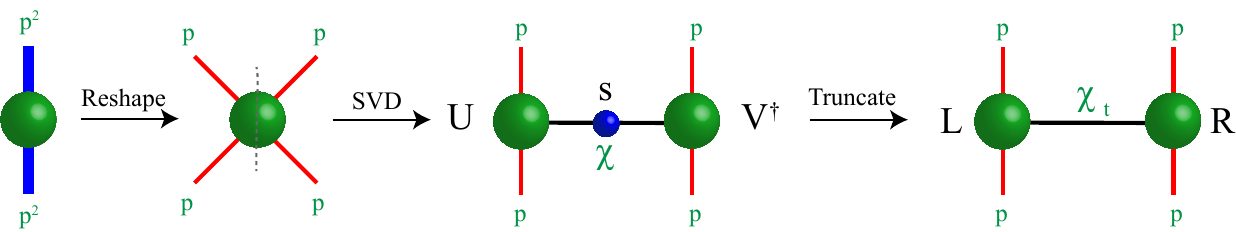}}
	\caption{[Color online] Singular value decomposition of a $p^2 \times p^2$ matrix $M$ into the $L$ and $R$ tensors of a 2-site MPO, see text for details.}
	\label{Fig:mpo_decompose}
\end{figure}

Large matrices and multi-dimensional arrays with a massive number of parameters can further be represented efficiently by tensor networks. Fig.~\ref{Fig:mps_mpo} shows examples of two well-known tensor networks, namely, matrix product states (MPS) (also known as a tensor trains) and matrix product operators (MPO) \cite{Verstraete2008}, which are used for representing one-dimensional quantum many-body wave functions and operators. In particular, one can provide efficient MPO representations of large matrices by reducing the number of parameters in a controlled way. More specifically, a matrix can be decomposed as an MPO by applying singular value decomposition (SVD) and truncating the negligible singular values. An example of MPO factorization for a matrix $M_{p^2 \times p^2}$ is shown in Fig.~\ref{Fig:mpo_decompose}. Reshaping $M$ into a rank-$4$ tensor and applying an appropriate SVD, one can represent matrix $M$ as a 2-site MPO with tensors $L_{pp\chi}$ and $R_{\chi p p}$:
\begin{equation}
M=U S V^\dagger = L R, \quad \quad L = U \sqrt{S}, \quad \quad R = \sqrt{S} V^\dagger,
\end{equation}
where $U$ and $V^\dagger$ are unitary matrices reshaped into rank-$3$ tensors and $S$ is a diagonal matrix of singular values. For the $p^2 \times p^2$ matrix $M$ there exist at most $\chi = p^2$ singular values. The non-zero singular values quantify the amount of entanglement (correlation) between the $L$ and $R$ MPO tensors. In a weakly correlated system, most of the singular values are close to zero and can be discarded so that only the $\chi_{t}\le\chi$ largest singular values are kept. One can, therefore, reduce the number of parameters along the so-called {\it virtual} tensor dimensions (black link) by keeping only the most relevant degrees of freedom for the correlations between $L$ and $R$. In turn, this also implies that the interconnecting tensor dimensions which emerge in the decomposition capture the relevant degrees of freedom quantifying correlations in the tensor network. 
 
\subsection{Tensorizing Standard Neural Networks}
\label{sec:tensorize-NN} 

\begin{figure}
\centerline{\includegraphics[width=14cm]{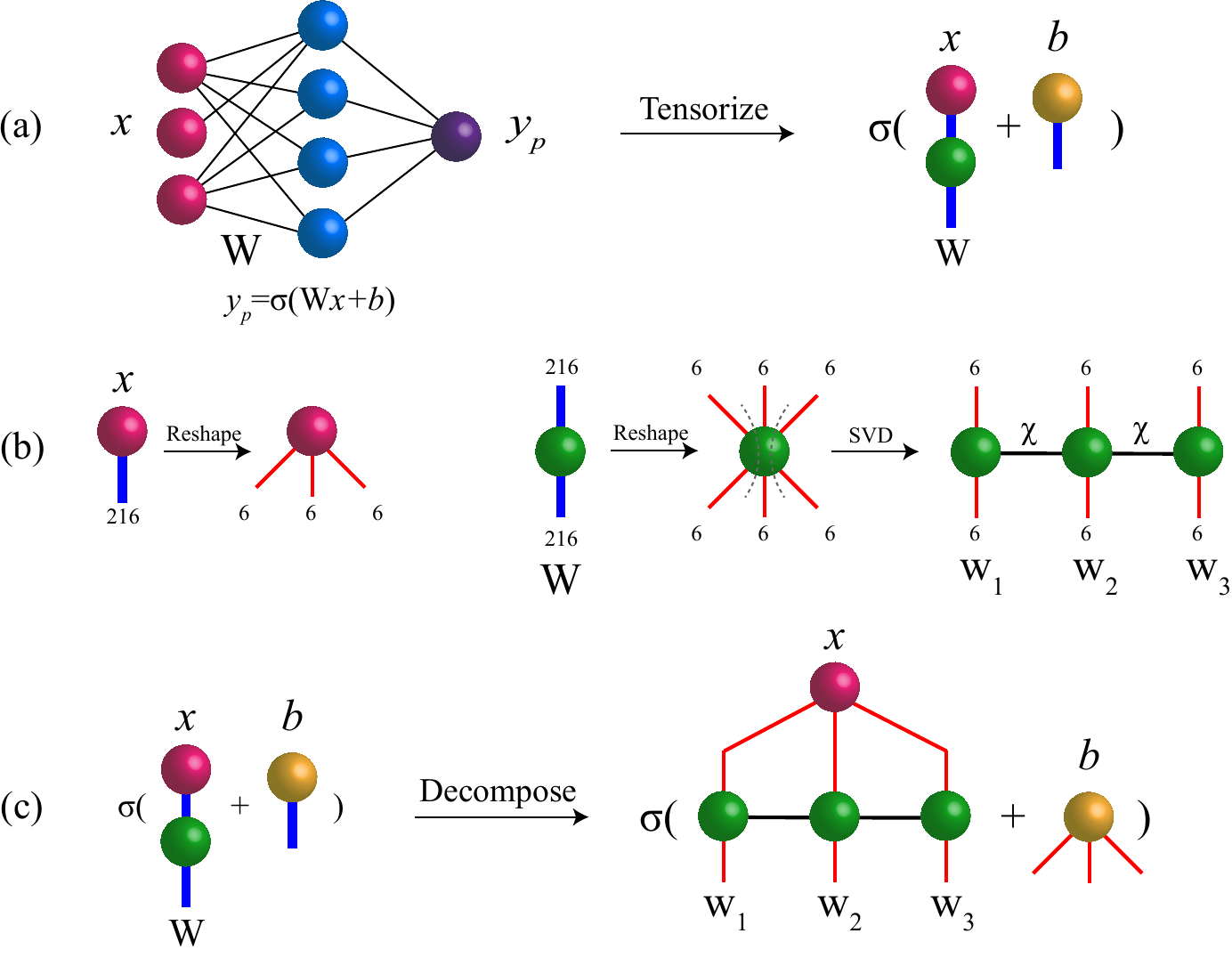}}
\caption{[Color online] Tensorizing a NN with one fully connected dense layer:  (a) TN representation of the NN; (b) MPO decomposition of the weight matrix $W$; (c) The resulting tensor neural network with MPO trainable weights.}
\label{Fig:tensorize_tnn}
\end{figure}

In this subsection, we show how to integrate tensor networks into standard neural networks to build a TNN. Fig.~\ref{Fig:tensorize_tnn}-(a) shows the structure of a classic NN with one fully connected \texttt{Dense} layer of hidden units (neurons). The network prediction, $y_p$, is obtained by feeding the input feature vector $x$ to the model as
\be
y_p = \sigma(Wx+b),
\label{eq:prediction}
\ee
where $W$ is the weight matrix, $b$ is some bias vector and $\sigma$ is the activation function (e.g., \texttt{ReLu} or \texttt{Sigmoid}). Training the NN amounts to finding the optimum values for the parameters of the $W$ weight matrix such that the $y_p$ corresponds the actual label of the data.

Depending on the size of the problem, the weight matrices can be considerably large. This introduces computational bottlenecks to the NN model from the point of view of the required memory for storing such matrices, and also due to the large training time. The problem becomes more dramatic when dealing with deep NNs with many hidden layers. Optimizing the parameters of such huge weights will reduce the accuracy and efficiency of the model. It is therefore a necessity to reduce the number of model parameters, without sacrificing accuracy. To this end, we should resort to a controllable truncation scheme such that we only discard the least important information and keep the most relevant one. An example of such clever data compression schemes based on TN and MPO decomposition has already been introduced in the previous subsection. Using MPO as an efficient representation for weight matrices of a NN was originally suggested in the ML community under the name tensor trains \cite{Novikov} and later reintroduced in other contexts for systematic compression of fully connected NN models \cite{Panagakis2021, Gao2020}, for solving partial differential equations with NNs \cite{Patel2022} and for language models \cite{Liu2021} and speech processing \cite{Sun2020}. {\color{blue} Applications of tensor trains for classification and regression has also been put forward in Ref.~\cite{qi_exploiting_2023}. Moreover, tensor networks have also been used for classical simulations of quantum circuits \cite{qi_theoretical_2023,chen_end_2021} which are also relevant for quantum machine learning and information processing.}   

Figure ~\ref{Fig:tensorize_tnn} further shows how the weight matrix of a fully connected \texttt{Dense} layer can be replaced by its MPO form obtained from consecutive applications of SVDs to the $W$ matrix. The new tensorized layer, called \texttt{TNLayer}, now has several trainable weights $w_i$ represented by MPOs. Reshaping the feature vector $x$ to match the MPO dimensions, the network prediction $y_p$ is obtained by contracting the resulting tensor network as shown in \ref{Fig:tensorize_tnn}-(c). The resulting tensorized neural network is then called a TNN.  

\begin{figure}
\centerline{\includegraphics[width=12cm]{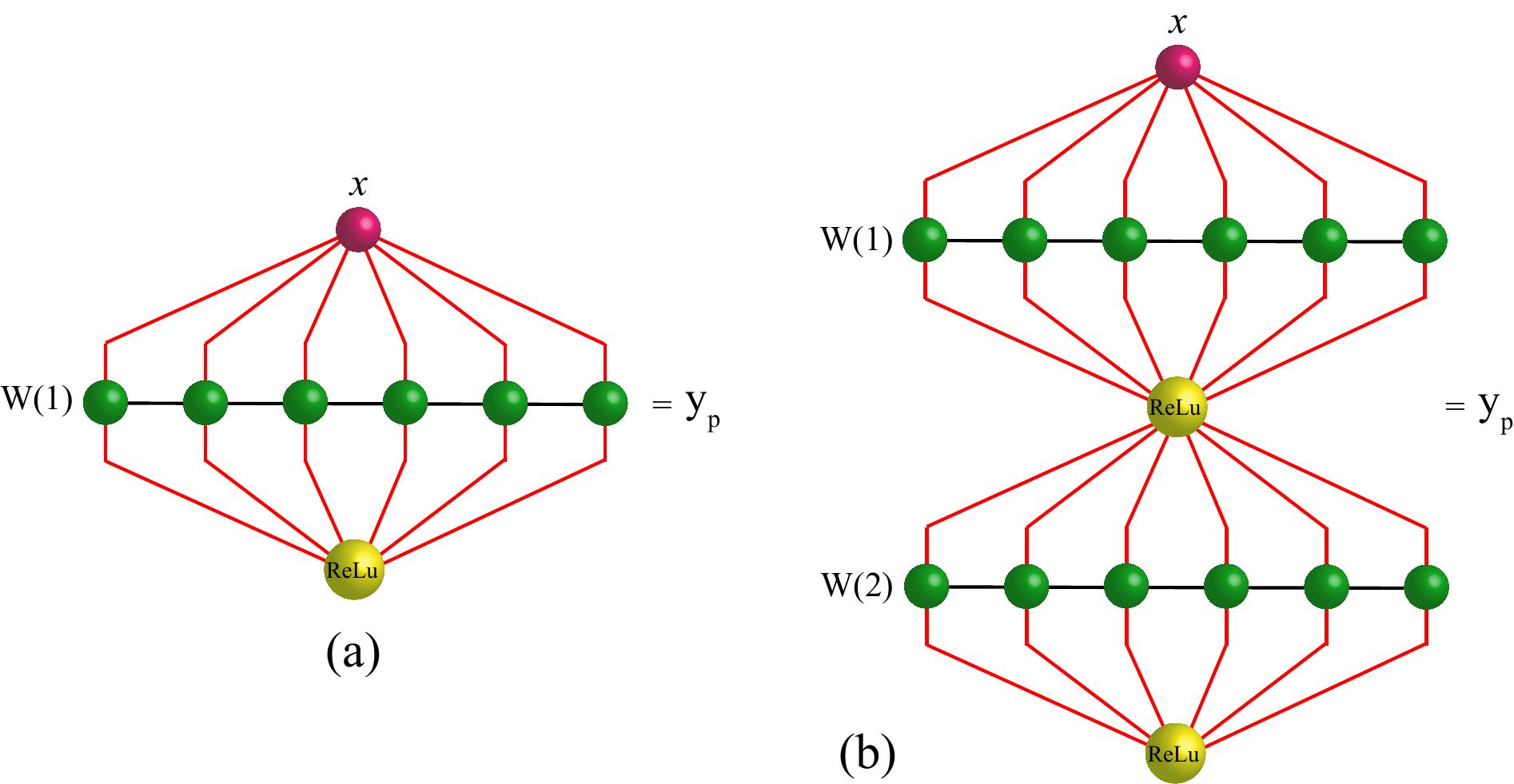}}
\caption{[Color online] Examples of fully tensorized TNNs with (a) one \texttt{TNLayer} and (b) with two \texttt{TNLayer}s.}
\label{Fig:example_tnn}
\end{figure}

While the idea of a TNN is generic, direct tensorization can be applied to models whose feature vector $x$ is factorable to integers, so that it could match the size and number dimensions of the MPO weights. An example of a valid-size feature vector $x$ with $216$ entries is shown in Fig.~\ref{Fig:tensorize_tnn}-(b). Such a vector can be reshaped into a rank-$3$ tensor with dimension size $6$ that is can be contracted to a \texttt{TNLayer} with three MPO weights, each with input size $6$. The features that are not factorizable to match the MPOs in the \texttt{TNLayer}, can rather be transformed in the preprocessing of the machine learning task so that their new size matches the input size of the \texttt{TNLayer}. Otherwise, one can introduce a dummy non-trainable \texttt{Dense} layer with size $N_F \times N_T$ in front of the \texttt{TNLayer} to compensate for the size mismatch. Here $N_F$ is the size of the feature vector and $N_T$ is the input size of the contracted  \texttt{TNLayer}. The non-trainable \texttt{Dense} layer can be initialized either randomly or according to a specific desired distribution. In theory, the choice of initialization for this dummy layer may slightly influence the convergence of the training algorithm and the path of the gradient toward the global minimum. In the experiments discussed in subsequent sections, we tested various random configurations and observed that their impact on the final outcomes was almost negligible.

Let us further stress that applying the activation on the TN layer is a non-linear operation and, therefore, cannot be tensorized or be applied to each MPO tensor individually. We, therefore, have to first contract the MPOs along their virtual legs to obtain a single weight matrix and then apply the activation on this matrix to obtain the network prediction, i.e.,
\be
y_p = \sigma({\rm tTR}[x_{ijk\ldots} w_i w_j w_k \ldots]+b_{ijk\ldots}),
\ee
where the subscripts $ijk\ldots$ run over the dimensions of the feature tensor $x$. Nevertheless, we showed this operation with a single activation tensor as illustrated in Fig.~\ref{Fig:example_tnn} for examples of TNN with one and two \texttt{TNLayer}s. In practice, one should first contract the features and MPOs, apply the activation function and then reshape the resulting tensor to match the inputs of the next layer. This process is continued until the whole network is contracted to obtain the prediction $y_p$. 

The mandatory contraction of MPO weights enforced by the activation function can in principle be useful because one can connect the contracted \texttt{TNLayer} to \texttt{Dense} layers and design hybrid TNN models for different ML tasks. As exemplified in Secs.~\ref{sec:regressor} and \ref{sec:classifier}, we will use combinations of \texttt{TNLayer} and \texttt{Dense} layers to build different TNN models for regression and classification.

\section*{DMRG-like Algorithm for TNNs}
\label{sec:dmrg}

In the previous section we introduced the tensor neural networks and how to build them from \texttt{TNLayer}s. In this section, we provide details about the DMRG-like training algorithm for multi-layer TNN models. Similar to generic optimizers that can be found in any ML library such as \texttt{TensorFlow} \cite{TensorFlow} or \texttt{PyTorch} \cite{PyTorch}, our DMRG-like algorithm is also of generic purpose and can be used for training different models ranging from regression to classification. 

\begin{figure}
\centerline{\includegraphics[width=10cm]{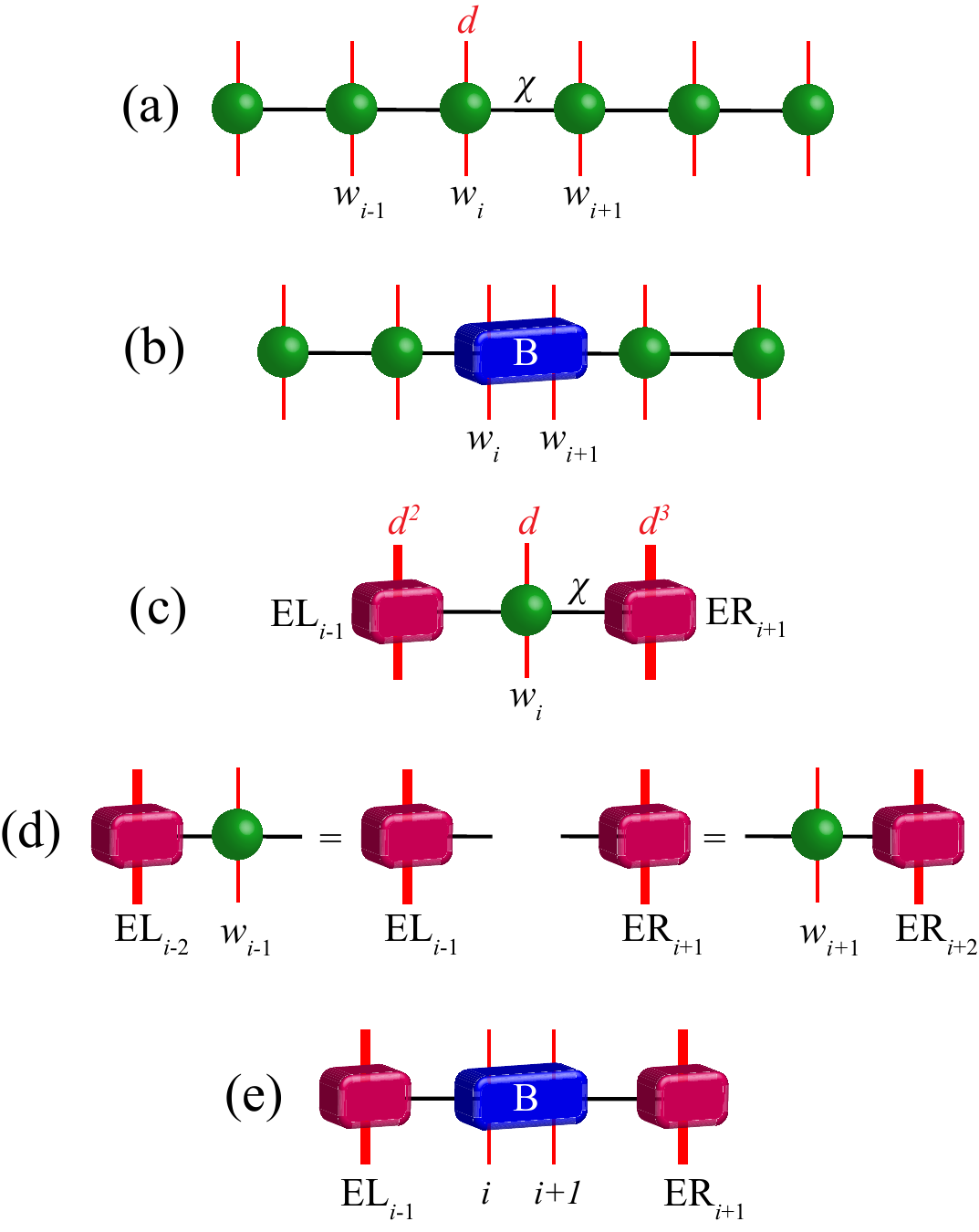}}
\caption{[Color online] Defining bond and environment tensors: (a) MPO factorized weights; (b) Bond tensor $B_{i,i+1}$ (shown in blue) which is obtained from contracting a pair of neighboring MPO tensors, $w_i, w_{i+1}$; (c) Left and right environment tensors of each $w_i$ MPO tensor; (d) A single step of obtaining environment tensors; (e) Alternative representation of MPO weights in terms of bond tensor $B_{i,i+1}$ and its surrounding environment.}
\label{Fig:tnn_weights_envs}
\end{figure}

\subsection{Local gradient-descent update} 
\label{sec:lgd}

The generic idea of training a neural network is to find the optimum values for the weight matrices of the NN layers by minimizing a loss function. In a feed-forward NN, all the data or batches of it are fed to the network to calculate the $y_p$ in the forward path. Next, the gradients of trainable weights with respect to the loss function are calculated in the backward path, and the weights are updated accordingly with a gradient-descent (GD) step. This whole process is iterated until a convergence criterion is met. Once trained, the model is tested to predict on unseen test data. The performance of the model is further evaluated by measuring different accuracy metrics. 

\begin{figure}
\centerline{\includegraphics[width=12cm]{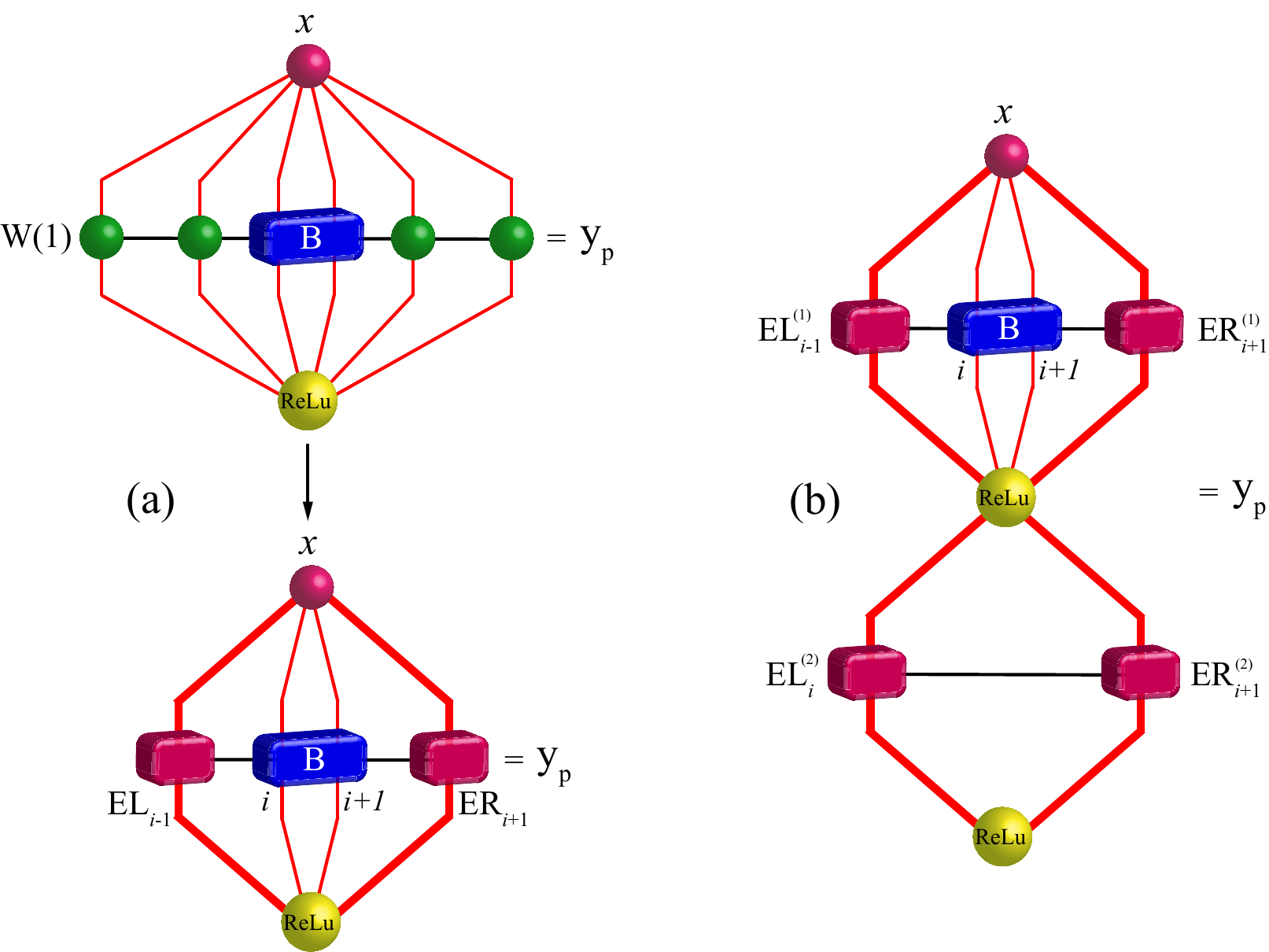}}
\caption{[Color online] Alternative representation of a TNN in terms of bond tensor $B_{i,i+1}$ and environments for (a) a TNN with one \texttt{TNLayer} and (b) two \texttt{TNLayer}s.}
\label{Fig:tnn_with_env}
\end{figure}

In contrast to classic NNs with \texttt{Dense} layers, the TNN is composed of \texttt{TNLayer}s that have multiple trainable MPO tensors. While a similar GD approach can be used for training the MPOs \cite{Patel2022}, here we use a DMRG-like technique to update the MPO tensors by pairs, with a sweeping local gradient-descent (LGD) algorithm. However, before we present the detail of the update, the following remarks are in order: i) Given a pair of neighboring MPO tensors, $w_i, w_{i+1}$, a {\it bond} tensor $B_{i,i+1}$ is obtained by contracting the $w_i$ and $w_{i+1}$ along their shared virtual dimension (see Fig.~\ref{Fig:tnn_weights_envs}-(a,b)). ii) For every MPO tensor $w_i$, the contraction of all tensors at the right side and left side of $w_i$ are called environment. These are denoted by $EL_{i-1}$, $ER_{i+1}$ for the left and right environments, respectively (see Fig.~\ref{Fig:tnn_weights_envs}-(c,d)). iii) For every bond tensor $B_{i,i+1}$, the \texttt{TNLayer} and further the overall TNN can be represented in terms of the bond tensor and its left and right environments, as shown in Fig.~\ref{Fig:tnn_weights_envs}-(e) and Fig.~\ref{Fig:tnn_with_env}. Given the fact that the TNN has to be contracted for every $B_{i,i+1}$ pair, which is required for updating the MPO weights, introducing the environment tensor can substantially reduce the computational cost of the network contraction.  

\begin{figure}
	\centerline{\includegraphics[width=11cm]{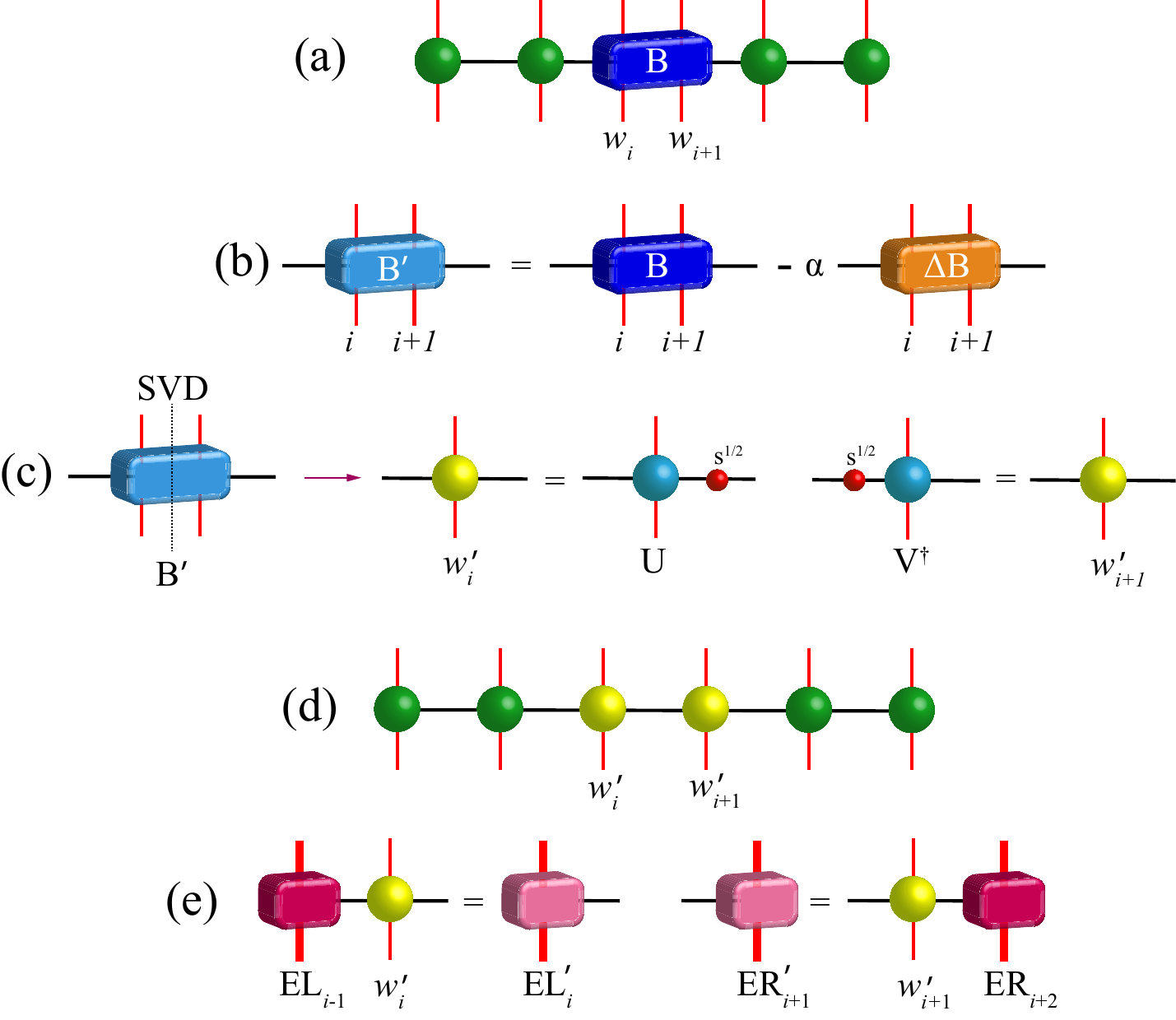}}
	\caption{[Color online] Graphical representation of the DMRG-like sweeping algorithm for training TNNs: (a) MPO weights with bond tensor $B_{i,i+1}$; (b) Updating the bond tensor with LGD step; (c) Calculating the updated local weights $w'_i$ and $w'_{i+1}$ from the updated bond tensor $B'_{i,i+1}$ by SVD and truncation; (e) Updating the left and right environments.}
	\label{Fig:local_gdu}
\end{figure}

Having the aforementioned remarks in mind, we train the TNN by sweeping over the MPO pairs of tensors of each \texttt{TNLayer} as follows:
\begin{enumerate}
	\item Do for all MPO tensor pairs $\{w_i, w_{i+1}\}$ where $i \in[1, N_{\rm MPO}-1]$ (left to right sweep): 
	\begin{enumerate}
		\item Build the local bond tensor $B_{i,i+1}$ by contracting $w_i$ and $w_{i+1}$. 
		\item Calculate the gradient of $B_{i,i+1}$, i.e., $\Delta B_{i,i+1}$ with respect to a loss function (see Sec.~\ref{sec:tensor-grad}, \ref{sec:auto-grad}). 
		\item Update the bond tensor with learning rate $\alpha$: $B'_{i,i+1} = B_{i,i+1} - \alpha  \Delta B_{i,i+1}$. 
		\item Split the updated bond tensor by SVD: $B'_{i,i+1} = U S V^{\dagger}$.
		\item Truncate the matrices to keep the desired number of singular values: $B'_{i,i+1} \approx U' S' V'^{\dagger}$.
		\item Update the weights: $w'_i = U' \sqrt{S'}$ and $w'_{i+1} = \sqrt{S'} V'^{\dagger}$.
		\item Update the left and right environment tensors: $EL'_{i}=EL_{i-1} w'_{i}$ and $ER'_{i+1}=w'_{i+1}ER_{i+2}$.
	\end{enumerate}
	\item Sweep back from right to left and repeat the same pair update.
\end{enumerate}
Details about the LGD algorithm in terms of TN diagrams are also provided in Fig.~\ref{Fig:local_gdu}.

Let us further note that other instances of DMRG-based training algorithms have already been introduced in previous studies such as in Ref.~\cite{Novikov} that the author uses a one-site sweeping algorithm for training the MPO layers or the single-layer DMRG sweeping algorithm of Ref.~\cite{Stoudenmire2016} for kernel learning. Our training algorithm differs from previous approaches in the sense that first, it is based on two-site DMRG sweeping update, which is known from simulation of quantum many-body systems to be more stable. The two site update, as will be discussed in future sections, can provide access to the entanglement structure of the weight tensors and shed light on the expressive power of TNNs. Second, in contrast to Ref.~\cite{Stoudenmire2016}, our sweeping algorithm is not limited to only a single layer and can be straightforwardly adapted to multilayer and deep networks with many layers. Moreover, our training algorithm allows designing TNN models with mixed \texttt{Dense} and \texttt{TNLayer}s, bringing huge flexibility to perform generic ML tasks.     

\subsection{Tensor gradient for linear activations} 
\label{sec:tensor-grad}

 \begin{figure}
	\centerline{\includegraphics[width=11cm]{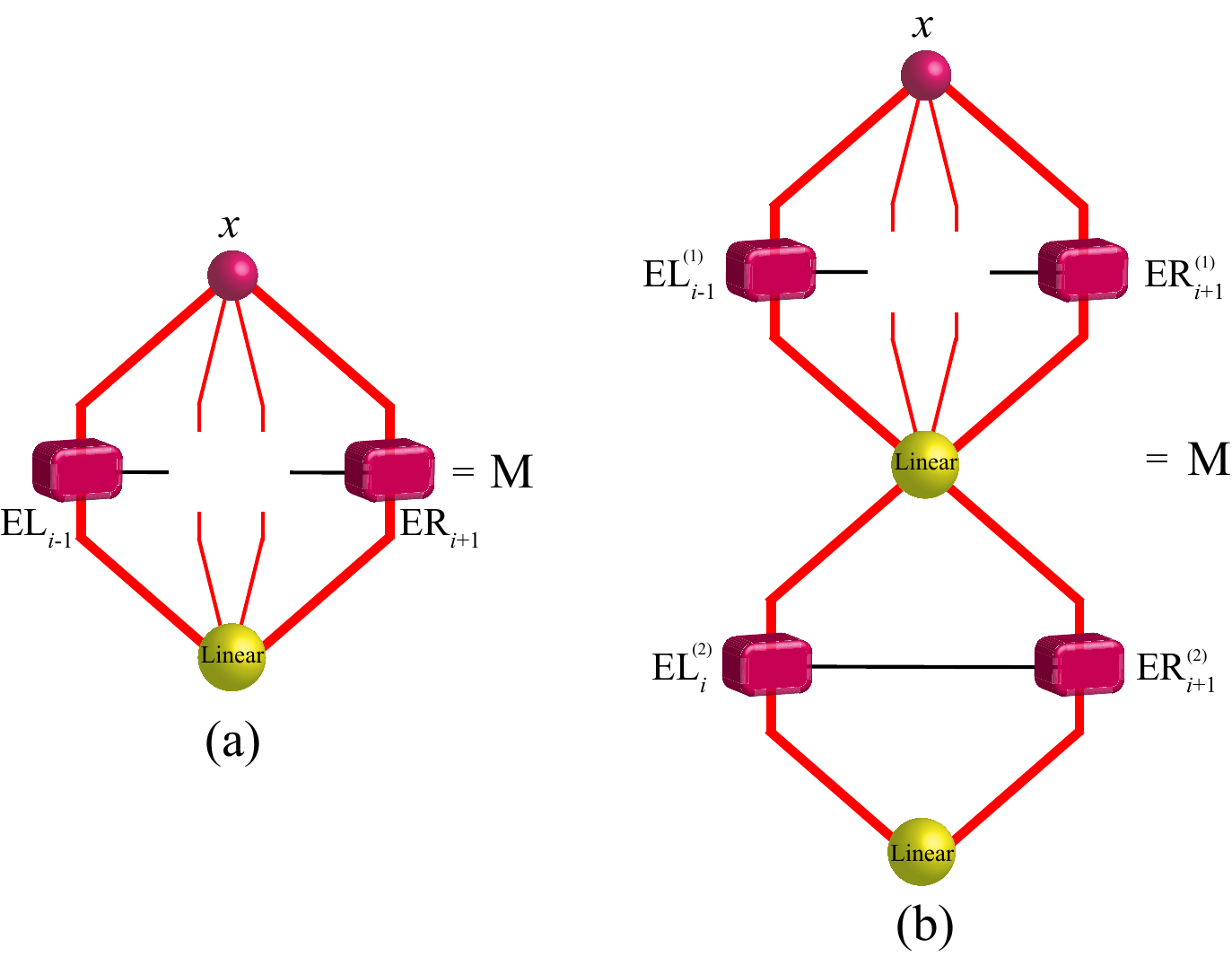}}
	\caption{[Color online] The $M$ tensor required for manual calculation of the gradient $\Delta B_{i,i+1}$ for a TNN with (a) a single \texttt{TNLayer}, and (b) two \texttt{TNLayer}s, and with linear activation functions.}
	\label{Fig:manual_gradient}
\end{figure}

The DMRG-like sweeping algorithm for training the TNN is a gradient-based optimization approach in which the local bond tensors, $B_{i,i+1}$, are updated towards a global minimum of a loss function by a gradient descent step with learning rate $\alpha$. Gradient of the bond tensors with respect to the loss, $\Delta B_{i,i+1}$, are therefore required for updating the weights. Given the fact that the TNN is a tensor network, gradients of the tensors in such a network with respect to a desired loss function are simply the network itself, with the corresponding tensor removed from the network. More specifically, consider a desired loss function such as the mean-squared error (MSE),
\be
\label{eq:mse}
L_{\rm MSE} = \frac{1}{N_s}\sum_{j=0}^{N_s}(y_p-y_j)^2 , 
\ee 
where $N_s$ is the number of training samples (input features) and $y_j$s are their labels. Defining $y_p=MB$, where $M$ is the contraction of all tensors in the TNN excluding the bond tensor $B$ (see Fig.~\ref{Fig:manual_gradient}), the $L_{\rm MSE}$ alternatively reads
\be
\label{eq:mse-MB}
L_{\rm MSE} = \frac{1}{N_s}\sum_{j=0}^{N_s}(MB - y_j)^2. 
\ee 
Taking the derivative with respect to $B$, the gradient of local bond tensors is given by 
\be
\label{eq:tensor-grad}
\Delta B=-\frac{\partial L_{\rm MSE}}{\partial B} =\frac{2}{N_s} \sum_{n=1}^{N_s} (y_j - y_p) M.
\ee
Note that while both the $M$ tensor and $y_p$ involve the contraction of the TNN, we do not need to do it twice. In practice, we first calculate the $M$ tensor and then we contract it to the bond tensor $B$ to obtain the $y_p$ prediction. 

The above procedure for manual calculation of the gradient is only valid for linear activations or any other activation function that can be applied to the individual MPO weight tensors and not the contracted network. The reason is that the $M$ tensor in Fig.~\ref{Fig:manual_gradient} is obtained by removing the bond tensor $B$ from the whole network after the application of the activation function. It has already been pointed out that for non-linear activations, one has to first contract the MPOs and the input from the previous layer and then apply the activation. It is obvious that the $B$ tensor can not be removed from a contracted network. Therefore, no $M$ tensor can be formed to calculate the tensor gradients manually. 

Let us further remark that while the gradient in Eq.~\eqref{eq:tensor-grad} and, subsequently, the LGD update (introduced in the previous subsection) are local, the DMRG-like sweep will restore the global features and correlations once iterated sufficiently.

\subsection{Automatic gradient} 
\label{sec:auto-grad}

Although the previous tensor gradient approach is suitable for TNN models fully composed of \texttt{TNLayer}s and linear activation functions, it will not be efficient and flexible once dealing with models with hybrid architectures containing a mixture of \texttt{Dense} and \texttt{TNLayer}s and nonlinear activation functions. Given the fact that our TNN is a feed-forward neural network, we can use automatic differentiation schemes and obtain the gradient of the TNN trainable weights with back-propagation (see Ref.~\cite{Margossian2019,Baydin2018} for a review on automatic differentiation and back-propagation). 

\begin{figure}
\centerline{\includegraphics[width=12cm]{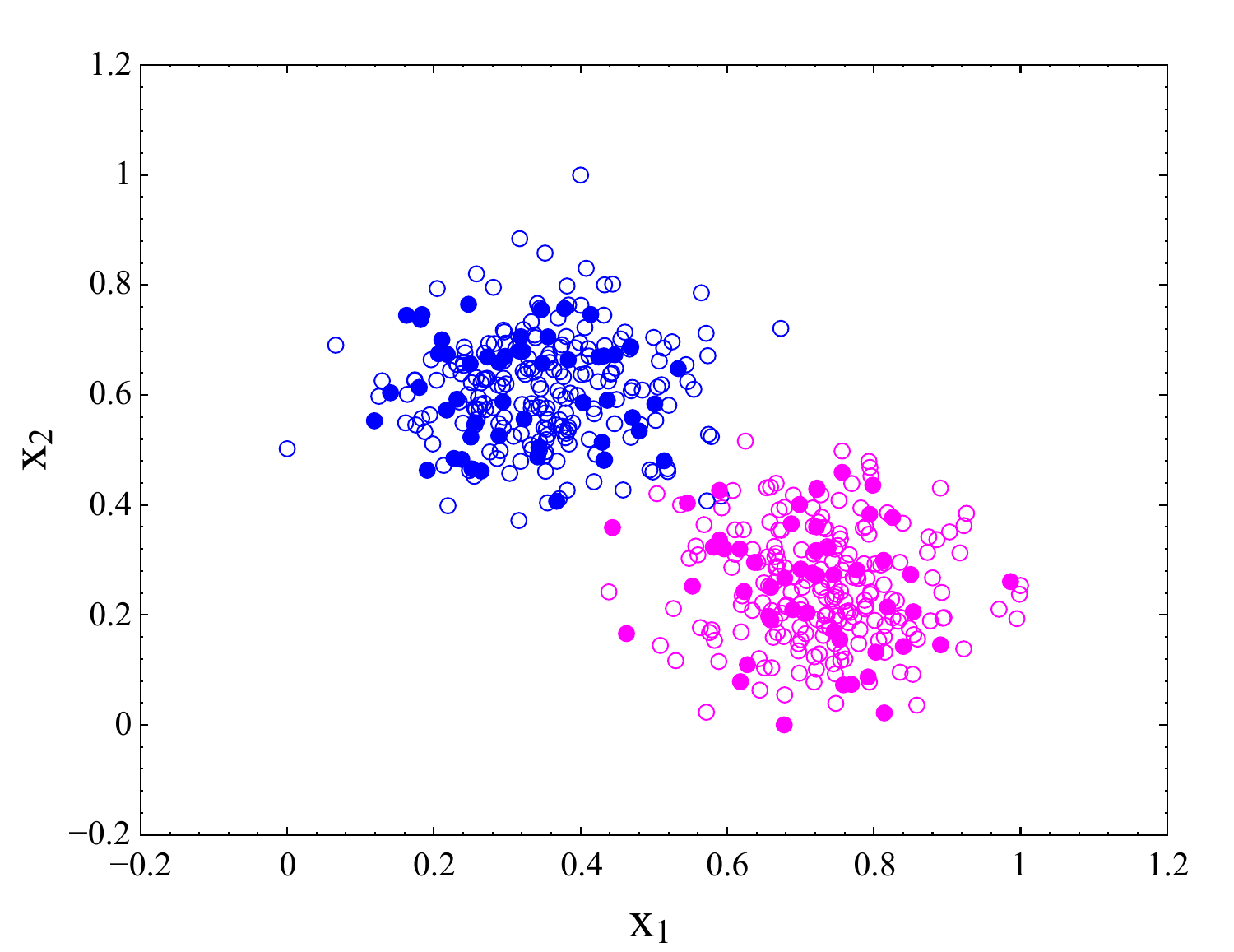}}
\caption{[Color online] Binary classification of isotropic Gaussian blobs with TNN \ref{model:blobs}. The empty (filled) circles represent training (test) data.}
\label{Fig:classificcation_blobs}
\end{figure}

While one can implement the whole process manually, implementing the \texttt{TNLayer} in one of the ML libraries that support the automatic differentiation is highly recommended. Here we used the \texttt{TensorFlow} library and one of its useful features, i.e., \texttt{GradientTape} \cite{TensorFlow} to record all the mathematical operations in the forward path. These include the contraction of all tensors and calculation of the network prediction, $y_p$, as well as the loss, ex. Eq.\eqref{eq:mse}. The \texttt{GradientTape} uses the operation records of the forward path and calculates the gradient of all trainable vectors, matrices and tensors in the backward move. Once the gradients are obtained, every trainable weights of the TNN can be updated. The MPOs are updated as prescribed in steps (c)-(g) of the DMRG-like algorithm of Sec.~\ref{sec:lgd}. The bias vectors, $b$, and the weight matrices of the \texttt{Dense} layers $W$ can further be updated as 
\be
\label{eq:dense-update}
W' = W - \alpha \Delta W, \quad\quad b' = b - \alpha \Delta b. 
\ee     
For hybrid TNN models composed of multiple \texttt{TNLayer}s and \texttt{Dense} layers one can in principle update the weights with different strategies, i.e., layer-by-layer (LbL) or partial-all-layer (PaL). In the LbL approach, we start updating from the first layer until we reach the last output layer. The \texttt{Dense} layers are updated according to Eq.~\eqref{eq:dense-update} and the MPO weights of \texttt{TNLayer}s are updated by a few sweeps say $N_{\rm sweep} \leq 10$ according to the DMRG-like algorithm of Sec.~\ref{sec:lgd}. On the other hand, the PaL approach targets all the layers at once. In this scheme, we consider a large $N_{\rm sweep} \geq 2000$ and update the MPO pairs $\{w^l_i, w^l_{i+1}\}$ of all \texttt{TNLayer}s ($l$ is the layer index), i.e, we do a multi-layer sweep. With each MPO pair update, we update the dense layers as well. This will help to reflect the local update of MPO weights in the matrix weights of the \texttt{Dense} layers during the sweep. The multi-layer sweep is performed until the loss converges  to a certain threshold or until the $N_{\rm sweep}$ is reached. 

It is also worth noting that for training the example models that we provide in the next sections for regression and classification, we observed that the PaL approach with large sweeps works much better from the point of view of convergence and accuracy. However the number of sweeps also depends on the dataset as well. For example for the case of MNIST handwritten digits since the number of record in the datasets are very large, even a single sweep is sufficient to obtain a high accuracy.

 \begin{table}
\begin{center}
\caption{TNN architecture for binary classification of random Gaussian blobs }
\label{model:blobs}
\begin{tabular}{l*{1}{c}r}
\hline 
$N_s = 500$, Sweep = $2000$, $\alpha = 0.1$, Batch = $500$ \\
\hline 
$\mathcal{I}=
\Big\{ \{ x_1, x_2 \}, \{ y \} \Big| \ 0 \leq x_1, x_2 \leq 1, \ y = [0,1]\Big\}$  
\\
$\downarrow
\ 
\left\{ \begin{array}{ll}
\texttt{Dense}\{\text{Units}=64, \text{trainable = } \texttt{False} \} & \\
\text{Activation: } \texttt{None } & \\
\end{array} \right.$
\\ \\
$\downarrow
\
\left\{ \begin{array}{ll}
\texttt{TNLayer}\{\text{Units} = 64,N_{\rm MPO}=6\} & \\
\text{Activation: } \texttt{ReLU } & 
\end{array} \right.$ 
\\ \\
$\downarrow
\ 
\left\{ \begin{array}{ll}
\texttt{Dense}\{\text{Units}=2 \} & \\
\text{Activation: } \texttt{Softmax } & \\
\end{array} \right.$
\\ 
$\mathcal{O} = \text{[0,1]}$
\\
$\text{Loss Function: } \texttt{BinaryCrossEntropy }$
\\
\hline  
\end{tabular}
\end{center}
\end{table}

\subsection{Entanglement Entropy}     
\label{sec:entropy}
In classic NN algorithms, the training convergence is checked by monitoring the loss function or some accuracy metric. The main drawback of these approaches is that they do not provide any information about the nature of the weights, the correlation among their parameters, and their expressive power. Thanks to the DMRG-like update, we have access to the singular values at every step of the sweep iterations, and this provides valuable information about the correlation (entanglement in quantum case) between MPO tensors. The individual singular values along the virtual dimensions of MPO tensors or their accumulative behavior in terms of the  entropy, $S = \sum_i \lambda_i^2 \log \lambda_i^2$, reveal the degree of correlation (entanglement) between pieces of trainable weights. While weakly correlated states are distinguished by zero or very small entanglement entropy with only a few non-zero singular values, entangled states will have non-vanishing $S$ and many non-zero singular values. Moreover, observing the behavior of entanglement entropy and singular values during the training can be used as another convergence measure, on top of the loss, which directly sheds light on the convergence of the MPO parameters.

Let us further point out that in each \texttt{TNLayer} with $N_{\rm MPO}$ tensors, we have $N_{\rm MPO}-1$ virtual bonds. We can calculate the entanglement entropy on each of these bonds. The entanglement entropy values may vary based on the model's architecture, the input data, and the size of the virtual dimensions. Therefore, one can examine the individual or average properties of these entanglement entropies to assess the convergence or correlations among the model parameters.  

 \begin{figure*}
	\centerline{\includegraphics[width=18cm]{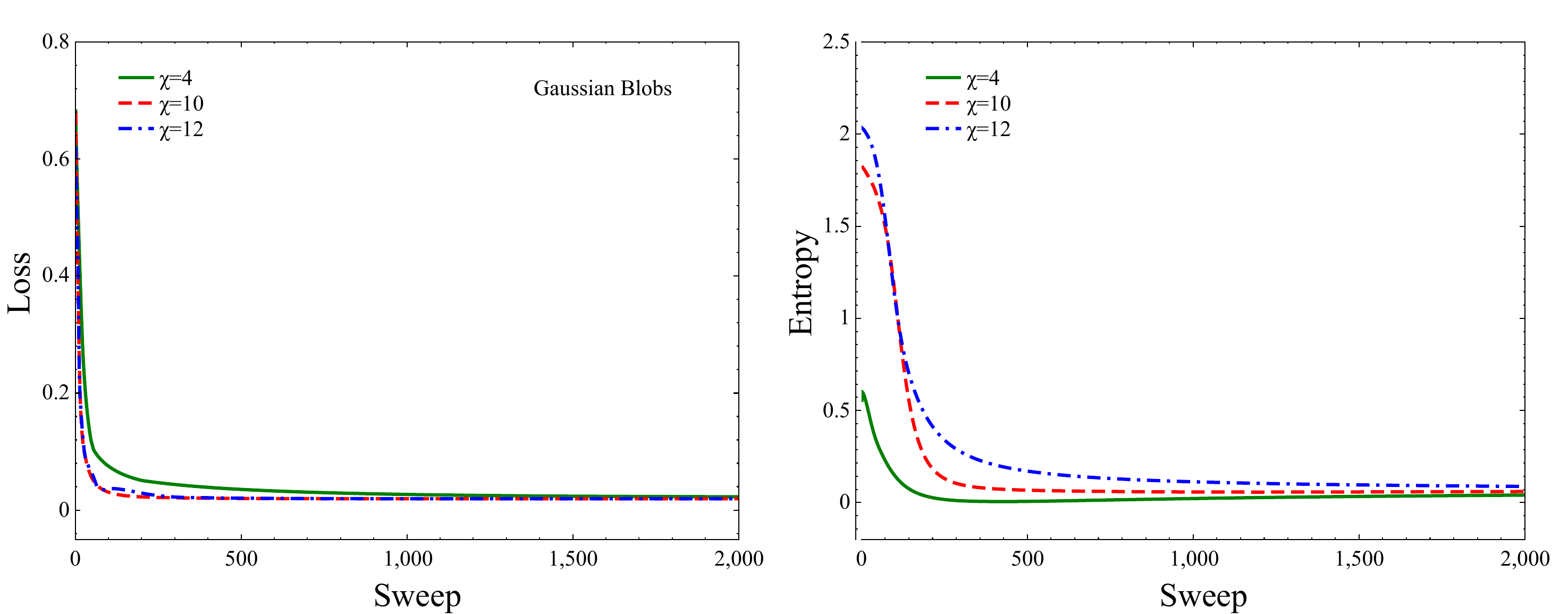}}
	\caption{[Color online] (left) Training loss and  (right) the $S$ entanglement entropy on the middle virtual bond of the MPO weights as a function of DMRG-like sweeps for training the TNN model \ref{model:blobs} used for classification of Gaussian blobs.}
	\label{Fig:loss_svn_blobs}
\end{figure*}

\section*{TNN Classifier}
\label{sec:classifier}
After introducing the TNN and the DMRG-like sweeping algorithm, let us see the performance and accuracy of the technique for classification tasks. In what follows, we present two examples for the classification of labeled data, one for the isotropic Gaussian blobs and another for the spiral distribution, and test the TNN in action.

\subsection{Gaussian Blobs} 
\label{sec:blobs}
As first example, we use the TNN to model a binary classification and train it over the random isotropic Gaussian blobs as shown with empty circles in Fig.~\ref{Fig:classificcation_blobs}. Each sample in the dataset has two position features denoted by $x_1, x_2$ and a label $y=[0,1]$. The architecture of the TNN model used for the binary classification of the Gaussian blobs is detailed in Table~\ref{model:blobs}. The model is composed of a \texttt{TNLayer} and two \texttt{Dense} layers. The first \texttt{Dense} is non-trainable and has only been added to compensate for the size mismatch between the features and the next \texttt{TNLayer} (see Sec.~\ref{sec:tensorize-NN}). This dummy layer has been initialized with random values. The \texttt{TNLayer} has six MPO trainable weights, each with input and output dimensions $d=2$, and virtual dimension $\chi$, and \texttt{ReLu} activation function. Finally the output layer is a \texttt{Dense} layer with \texttt{Softmax} activation which delivers the predicted probabilities of both labels as one-hot-encoded (OHE) vectors. The prediction $y_p$ can then be read form the \texttt{argmax} of the OHE probabilities for each sample.

Fig.~\ref{Fig:classificcation_blobs} shows the distribution of the two Gaussian blobs clusters. The empty circles denote the training data and the filled circles represent the test data. We performed the training for virtual dimension $\chi \in[2,12]$ over $500$ training data and tested the model over $100$ unseen data. The average accuracy for classifying Gaussian blobs with the TNN model \ref{model:blobs} is $99\%$ (see Table \ref{model:blobs} for training hyper-parameters).

Fig.~\ref{Fig:loss_svn_blobs} further shows the loss function and entanglement entropy on the middle virtual bond of the MPO weights, as a function of the sweep iterations for different bond dimensions $\chi$. While both plots provide a measure for the training convergence, the $S$ suggests that the MPO weights of the \texttt{TNLayer} represent a weekly entangled state with a small correlation among the parameters. This is best confirmed by very small entanglement entropy (close to zero) which is a typical behavior for product states. Looking at the distribution of blobs that are categorized into two localized clusters with almost no overlap, it is expected of the MPO weights be very close to that of the product states. The TNN therefore, not only classifies the blobs but also provides expressive insight into the nature of MPO weights. 

\subsection{Spiral Distribution} 
\label{sec:spiral}

\begin{figure}
\centerline{\includegraphics[width=12cm]{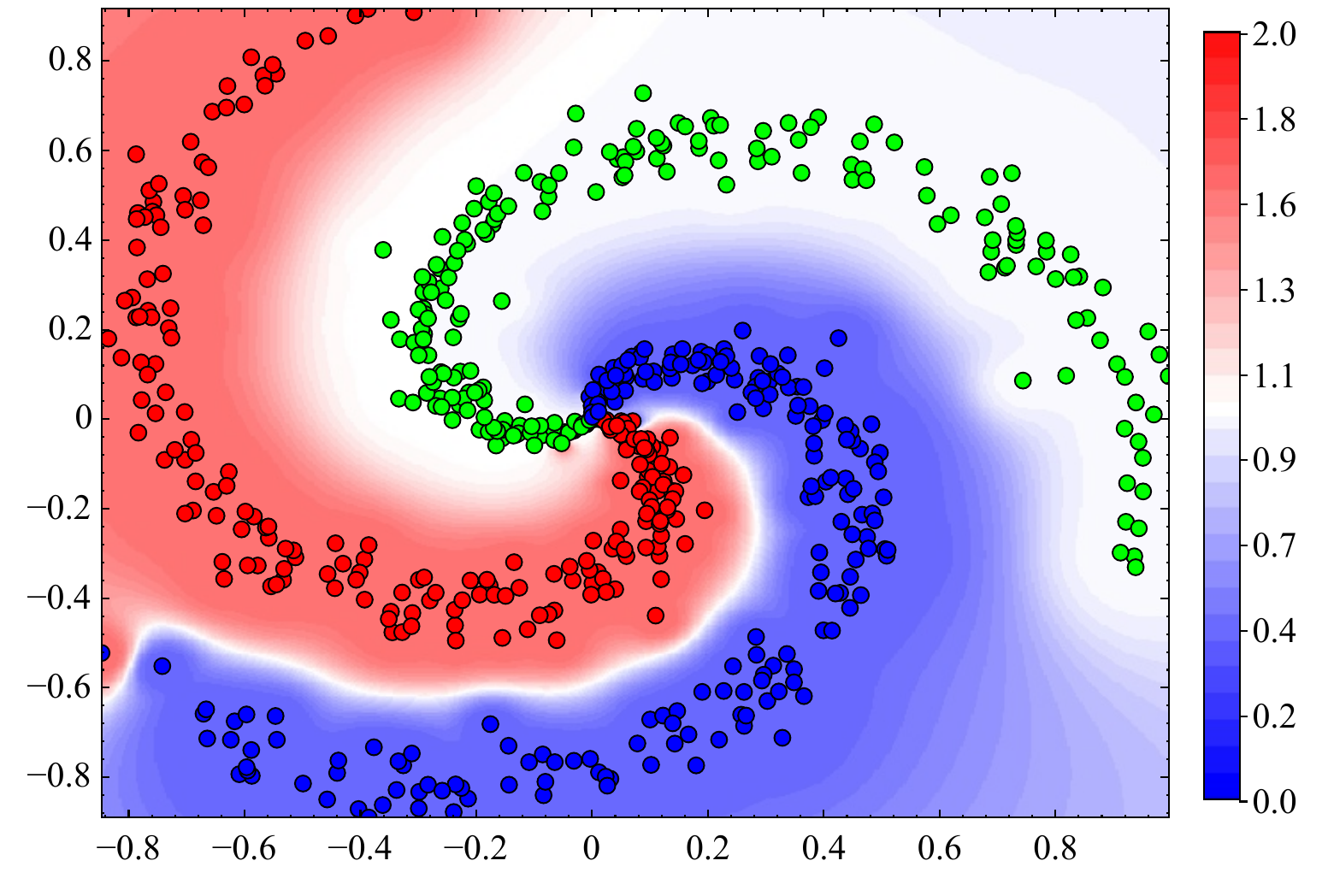}}
\caption{[Color online] Ternary classification of random spiral distribution with three classes. The red, blue and green filled circles distinguish the training data in each class. The red, white and blue shaded regions further distinguish the decision boundary obtained from training the TNN model \ref{model:spiral}.}
\label{Fig:spiral}
\end{figure}

\begin{table}
\begin{center}
\caption{TNN architecture for the classification of random spiral distributions}
\label{model:spiral}
\begin{tabular}{l*{1}{c}r}
\hline 
$N_s = 600$, Sweep = $3000$, $\alpha = 0.1$, Batch = $600$ \\
\hline 
$\mathcal{I}=
\Big\{ \{ x_1, x_2 \}, \{ y \} \Big| \ -1 \leq x_1, x_2 \leq 1, \ y = [0,1,2] \Big\}$  
\\
$\downarrow
\ 
\left\{ \begin{array}{ll}
\texttt{Dense}\{\text{Units}=64, \text{trainable = } \texttt{False} \} & \\
\text{Activation: } \texttt{None } & \\
\end{array} \right.$
\\ \\
$\downarrow
\
\left\{ \begin{array}{ll}
\texttt{TNLayer}\{\text{Units} = 64,N_{\rm MPO}=6\} & \\
\text{Activation: } \texttt{ReLU } & 
\end{array} \right.$ 
\\ \\
$\downarrow
\ 
\left\{ \begin{array}{ll}
\texttt{Dense}\{\text{Units}=3 \} & \\
\text{Activation: } \texttt{Softmax } & \\
\end{array} \right.$
\\ 
$\mathcal{O} = \text{[0,1,2]}$
\\
$\text{Loss Function: } \texttt{BinaryCrossEntropy }$
\\
\hline  
\end{tabular}
\end{center}
\end{table}

For the second example, we consider a random spiral distribution with three classes as illustrated in Fig.~\ref{Fig:spiral}. We use the TNN model described in Table \ref{model:spiral} for the ternary classification of spiral data. The model is similar to the previous example except that now the output \texttt{Dense} layer has three outputs for the probabilities of each class. We trained the model over $600$ samples, $200$, for each class and obtained the decision boundary of the spiral distribution as depicted in the shaded regions with red, white, and blue colors in Fig.~\ref{Fig:spiral}. The average accuracy of the model from different runs is $95\%$ (see Table \ref{model:spiral} for training hyper-parameters).

\begin{figure*}
	\centerline{\includegraphics[width=18cm]{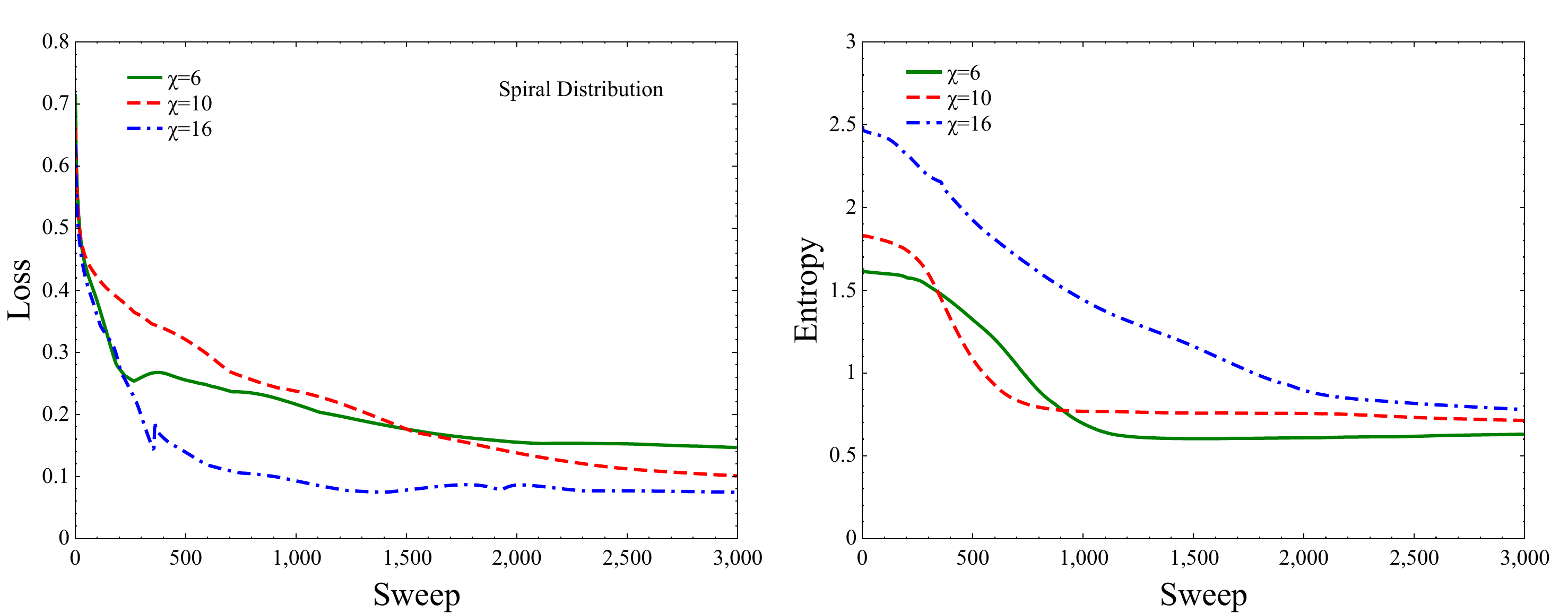}}
	\caption{[Color online] (left) Training loss and (right) the $S$ entanglement entropy on the middle virtual bond of the MPO weights as a function of DMRG-like sweeps for training the TNN model \ref{model:spiral} used for classification of spiral distibution.}
	\label{Fig:loss_svn_spiral}
\end{figure*}

The loss function and entanglement entropy, on the middle virtual bond of the MPO weights, during the training have also been displayed in Fig.~\ref{Fig:loss_svn_spiral} for different bond dimensions. Compared to gaussian blobs, one can clearly see that the behavior of loss and entanglement is totally different for the spiral distribution. Looking at the distribution of points in Fig.~\ref{Fig:spiral}, the spiral dataset is expected to be more correlated and therefore, more challenging to be trained. The behavior of the loss as well as the non-vanishing entanglement entropy indeed confirms the higher degree of entanglement among the parameters of the MPOs for the spiral distribution. The $S$ for the spiral distribution is approaching $\approx 0.75$ signaling an entangled structure with more correlated parameters than that of the Gaussian blobs.   

\section*{TNN Regressor} 
\label{sec:regressor}
In this section, we challenge our TNN for regression tasks. Here we present two regression examples one for fitting a line to a random linear distribution and another one for fitting to the random points around the nonlinear sine function.     

\begin{figure}
\centerline{\includegraphics[width=12cm]{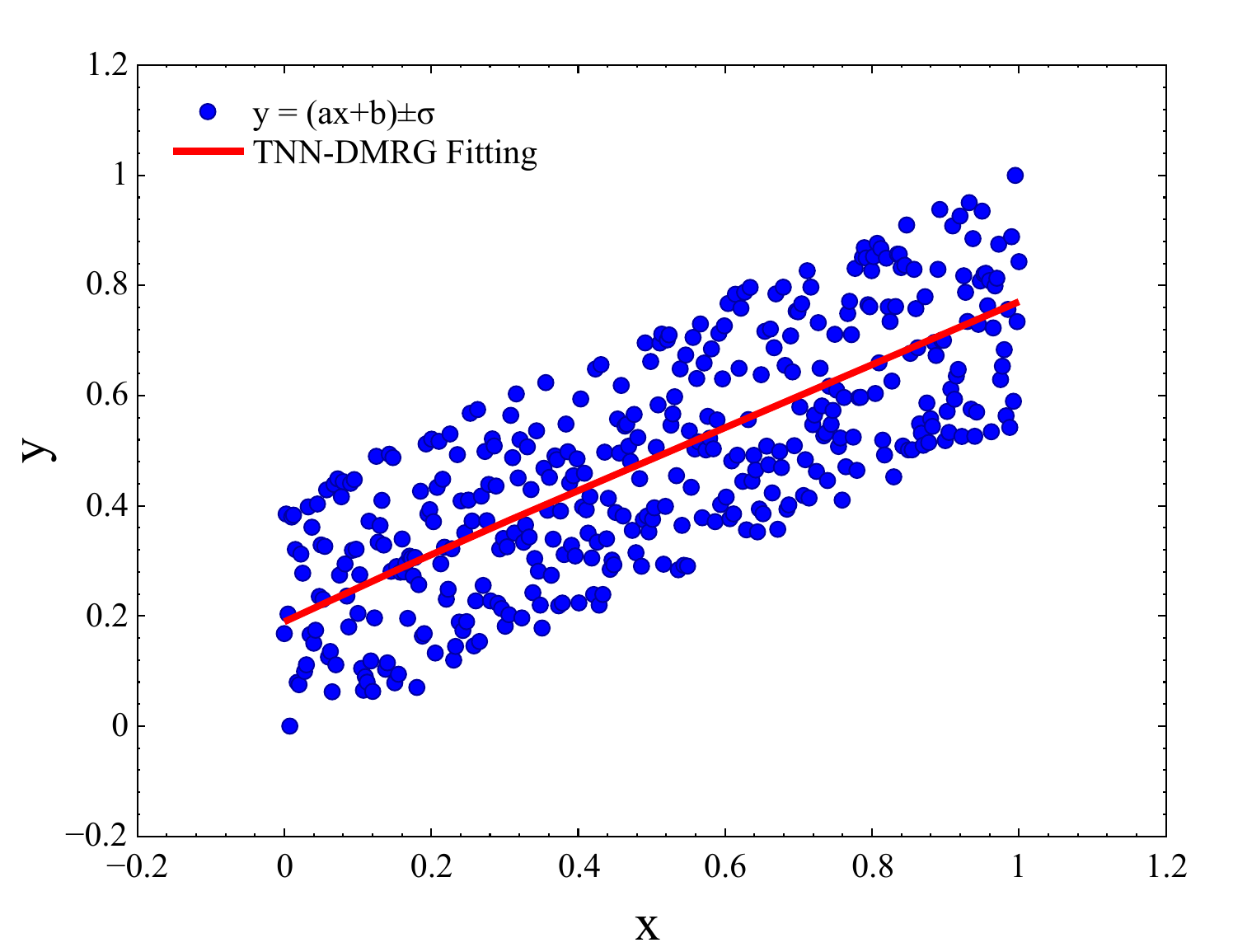}}
\caption{[Color online] Linear regression of the random data around the line $y=ax+b$ with $a=0.55$ and $b=0.20$. The red line is the fit obtained from training the TNN model \ref{model:linear}. The predicted slope and shift of the red fitted line are $a=0.57$ and $b=0.19$.}
\label{Fig:linear-regression}
\end{figure}

\subsection{Linear Regression} 
\label{sec:linear-reg}

\begin{table}
\begin{center}
\caption{TNN architecture for linear regression}
\label{model:linear}
\begin{tabular}{l*{1}{c}r}
\hline 
$N_s = 400$, Sweep = $2000$, $\alpha = 0.1$, Batch = $400$ \\
\hline 
$\mathcal{I}=
\Big\{ \{ x \}, \{ y \} \Big| \ 0 \leq x,y \leq 1 \Big\}$  
\\
$\downarrow
\ 
\left\{ \begin{array}{ll}
\texttt{Dense}\{\text{Units}=64, \text{trainable = } \texttt{False} \} & \\
\text{Activation: } \texttt{None } & \\
\end{array} \right.$
\\ \\
$\downarrow
\
\left\{ \begin{array}{ll}
\texttt{TNLayer}\{\text{Units} = 64,N_{\rm MPO}=6\} & \\
\text{Activation: } \texttt{ReLU } & 
\end{array} \right.$ 
\\ \\
$\downarrow
\ 
\left\{ \begin{array}{ll}
\texttt{Dense}\{\text{Units}=1 \} & \\
\text{Activation: } \texttt{None } & \\
\end{array} \right.$
\\ 
$\mathcal{O} = y_p$
\\
$\text{Loss Function: } \texttt{MeanSquareError}$
\\
\hline  
\end{tabular}
\end{center}
\end{table}

In order to test the TNN for a linear regression problem, we generate our data by adding uniform random noise to the points out of linear function $y=ax+b$ with ($a=0.55, b=0.20$)for $0\leq x \leq 1$. The distribution of the linear random points has been depicted with blue circles in Fig.~\ref{Fig:linear-regression}. In order to fit a line to the points, we designed a linear TTN model as described in Table \ref{model:linear}. The model is composed of a single trainable \texttt{TNLayer} with \texttt{ReLu} activation and a \texttt{Dense} output layer with no activation. As usual, we put a random non-trainable \texttt{Dense} layer in front of the \texttt{TNLayer} to compensate for the shape mismatch between the features and the \texttt{TNLayer}. 

Training the linear TNN model for $400$ random points and $2000$ sweeps, we obtain a linear line that is perfectly fitted to the random data as shown in red in Fig.~\ref{Fig:linear-regression}. Reading the slope and data shift from the predicted fitted line, we obtain $a_p=0.57$ and $b_p=0.19$ which is in good agreement with the original $a$ and $b$ parameters.     

\subsection{Non-linearRegression} 
\label{sec:nonlinear-reg}

Lastly, we challenge our TNN and DMRG-like algorithm for a non-linear regression task to predict a fit to the non-linear random points obtained by uniform noise to the $y=\sin x$ function for $0\leq x \leq 2\pi$. To this end, we introduce a non-linear TNN model with three \texttt{TNLayer}s as detailed in Table \ref{model:non-linear}. Neglecting the initial dummy \texttt{Dense} layer, the first \texttt{TNLayer} has a \texttt{Linear} activation, and the two others have \texttt{Sigmoid} activation functions. Adding to this structure an output \texttt{Dense} layer with \texttt{ReLu} activation, the resulting TNN model is fully capable of capturing non-linear correlation of the input features.

\begin{figure}
\centerline{\includegraphics[width=12cm]{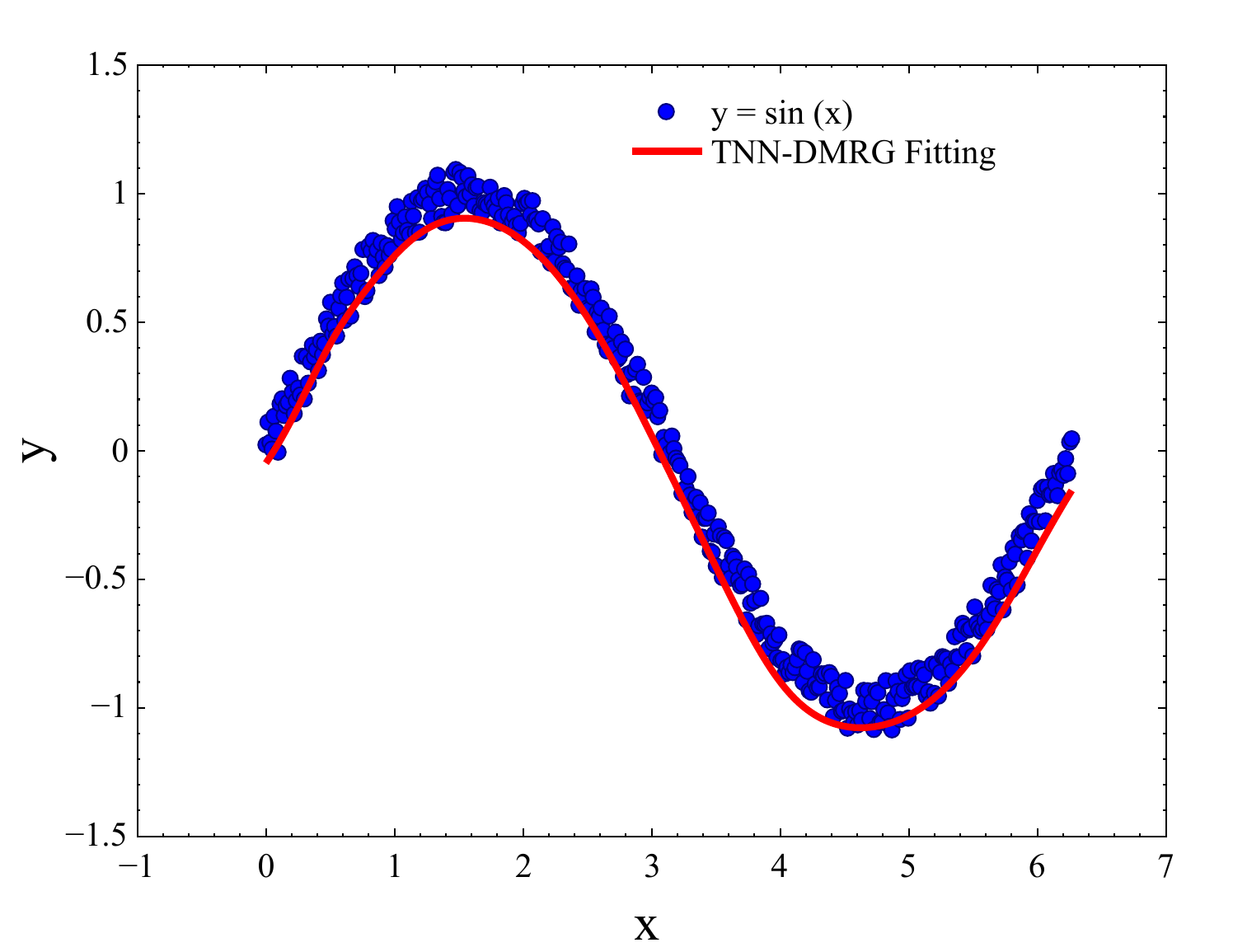}}
\caption{[Color online] Non-Linear regression of the random data introduced to the function $y=\sin x$ for $0\leq x \leq 2\pi$. The red line is the fit obtained from training the TNN model \ref{model:non-linear}.}
\label{Fig:non-linear-regression}
\end{figure}

Fig.~\ref{Fig:non-linear-regression} shows $400$ non-linear random $sine$ shape points that have been fitted perfectly by the predicted red line from the trained TNN model. The TNN architecture of model \ref{model:non-linear} is an example of a multi-layer deep TNN which is a mixture of both \texttt{Dense} and \texttt{TNLayer}s. Indeed other complicated architectures with more layers and different activations can be designed for generic ML tasks based on neural networks. The choice of these examples and hyper-parameters was to showcase the performance, accuracy, and flexibility of designing generic models with the TNN and DMRG sweeping algorithm. Indeed by changing the architecture or hyper-parameter tuning of the aforementioned models, one can improve the quality of the results. However, this was not the purpose of this study.

\begin{table}
\begin{center}
\caption{TNN architecture for non-linear regression}
\label{model:non-linear}
\begin{tabular}{l*{1}{c}r}
\hline 
$N_s = 400$, Sweep = $2000$, $\alpha = 0.1$, Batch = $400$ \\
\hline 
$\mathcal{I}=
\Big\{ \{ x \}, \{ y \} \Big| \ 0 \leq x \leq 2\pi, \ -1 \leq y \leq 1 \Big\}$  
\\
$\downarrow
\ 
\left\{ \begin{array}{ll}
\texttt{Dense}\{\text{Units}=64, \text{trainable = } \texttt{False} \} & \\
\text{Activation: } \texttt{None } & \\
\end{array} \right.$
\\ \\
$\downarrow
\
\left\{ \begin{array}{ll}
\texttt{TNLayer}\{\text{Units} = 64,N_{\rm MPO}=6\} & \\
\text{Activation: } \texttt{Linear } & 
\end{array} \right.$ 
\\ \\
$\downarrow
\
\left\{ \begin{array}{ll}
\texttt{TNLayer}\{\text{Units} = 64,N_{\rm MPO}=6\} & \\
\text{Activation: } \texttt{Sigmoid } & 
\end{array} \right.$ 
\\ \\
$\downarrow
\
\left\{ \begin{array}{ll}
\texttt{TNLayer}\{\text{Units} = 64,N_{\rm MPO}=6\} & \\
\text{Activation: } \texttt{Sigmoid } & 
\end{array} \right.$ 
\\ \\
$\downarrow
\ 
\left\{ \begin{array}{ll}
\texttt{Dense}\{\text{Units}=1 \} & \\
\text{Activation: } \texttt{ReLu } & \\
\end{array} \right.$
\\ 
$\mathcal{O} = y_p$
\\
$\text{Loss Function: } \texttt{MeanSquareError}$
\\
\hline  
\end{tabular}
\end{center}
\end{table}

\section*{TNN Image Recognition} 
\label{sec:mnist}

\begin{figure}
\centerline{\includegraphics[width=12cm]{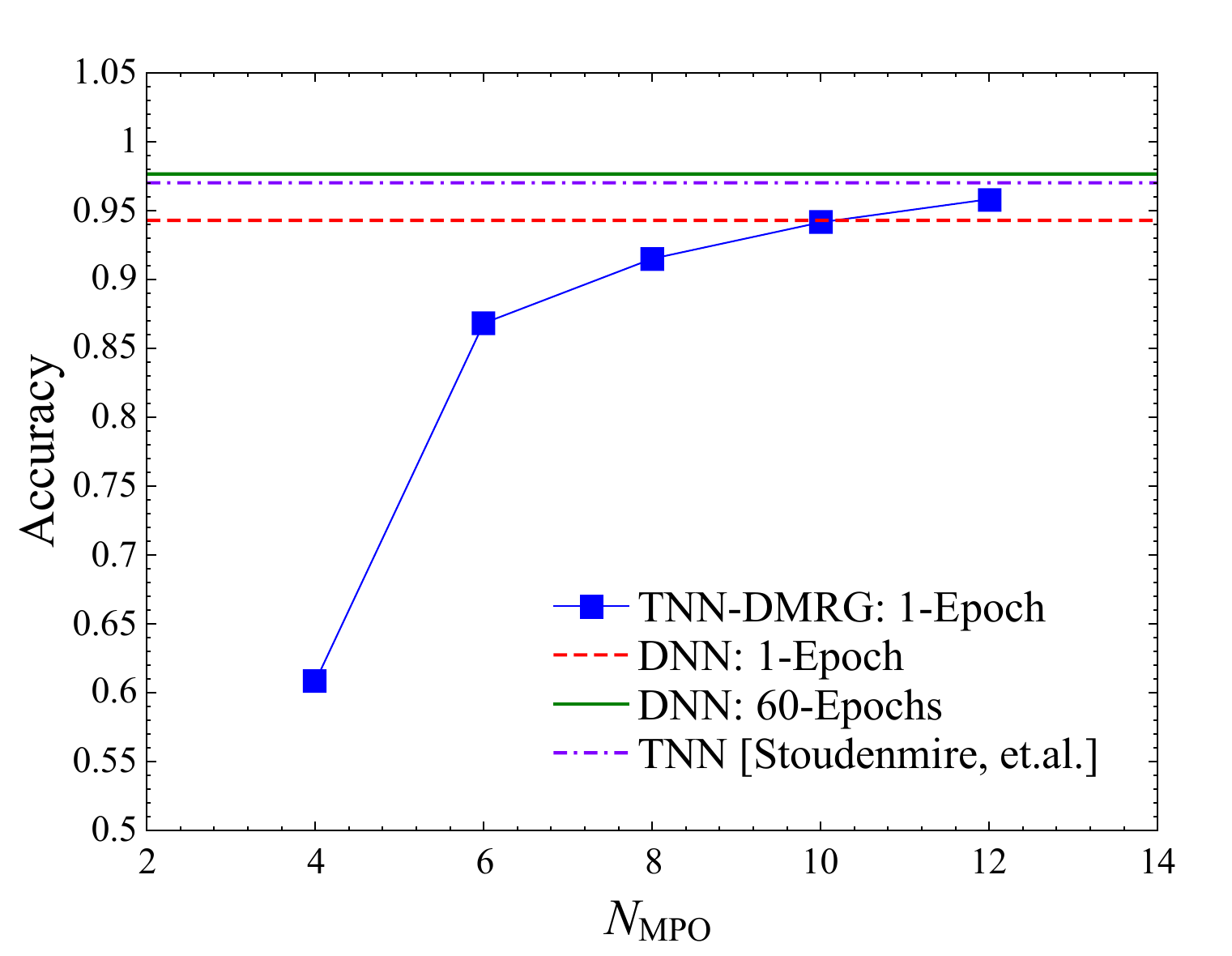}}
\caption{[Color online] Scaling of the MNIST classification accuracy with the number of tensors, $N_{\rm MPO}$, in the \texttt{TNLayer} of the TNN model for one training epoch. The red-dashed and solid-green lines represent the accuracy of the classic NN model for $1$ and $60$ epochs (full convergence), respectively. {\color{blue} The violet dashed-dot line further shows the accuracy of the TNN approach from Ref.~\cite{Stoudenmire2016}.}}
\label{Fig:mnist-accuracy-scaling}
\end{figure}

Next, to evaluate our algorithm's effectiveness in a more complex scenario, we applied the TNN model and DMRG-like sweeping algorithm to image recognition using the MNIST dataset of handwritten digits. We compared the results to those of a classical neural network (NN) model. To this end, we designed a TNN model featuring a single  \texttt{TNLayer} with \texttt{Relu} activation followed by an output \texttt{Dense} layer equipped with $10$ units to predict the probabilities of the digits zero through nine.  As usual, we introduced a dummy \texttt{Dense} layer with random configuration in front of the \texttt{TNLayer} to compensate for the shape mismatch with the input data. In contrast, the classical NN model employed a similar architecture but substituted the \texttt{TNLayer} with a standard \texttt{Dense} layer and omitted the dummy input layer.

While we used the DMRG-like algorithm for training the TNN model with learning rate $\alpha=0.60$, the classical NN model was trained using the built-in \texttt{Adam} optimizer of the \texttt{TensorFlow} library with learning rate $\alpha=0.001$. We used \texttt{BinaryCrossentropy} loss function for both models and benchmarked the accuracy of the two models in different scenarios. The MNIST dataset contains $60,000$ ($10,000$) training (test) dataset that we used all of them for training and testing the accuracy of both models. Besides, we trained the models on a core-$i9$ MacBook Pro 2019 with six $2.9$ GHz CPUs. This is relevant later where we benchmark the training time of both models.

\begin{figure}
\centerline{\includegraphics[width=18cm]{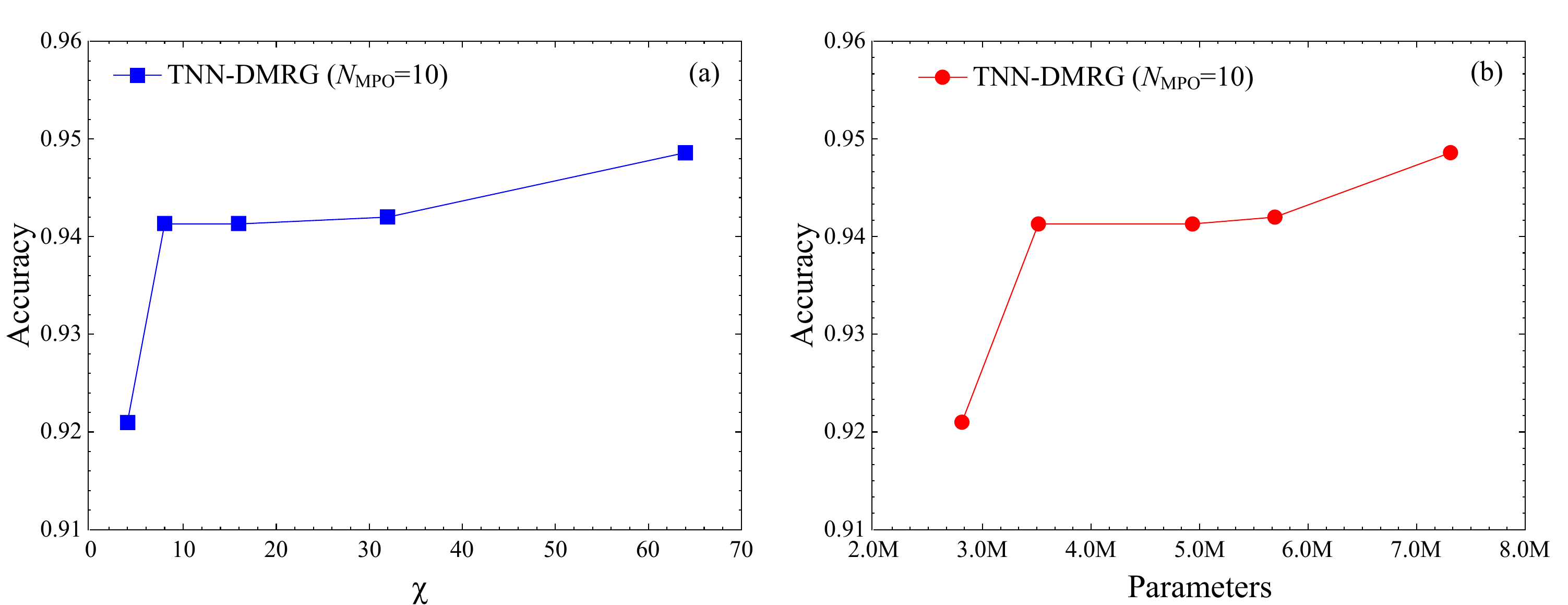}}
\caption{[Color online] Scaling of the MNIST classification accuracy (a) versus MPO virtual dimension $chi$ in the \texttt{TNLayer} and (b) versus the total number of trainable parameters of the TNN model, for one training epoch. The accuracies were obtained for fixed number of tensors, i.e, $N_{\rm MPO}=10$.}
\label{Fig:mnist-chi-scaling}
\end{figure}

Fig.~\ref{Fig:mnist-accuracy-scaling} illustrates the test accuracy of the TNN model as a function of the number of tensors,$N_{\rm MPO}$, in the \texttt{TNLayer}. The graph demonstrates that the accuracy in recognizing handwritten digits improves with an increase in the number of MPO tensors. The accuracies were recorded after one epoch and one sweep. Among these measurements, the highest accuracy observed was $0.958$ at $N_{\rm MPO}=12$. In contrast, the accuracy of the corresponding classical NN model was $0.942$ after one epoch, as indicated by the red-dashed line in the figure. {\color{blue} For the reference, the accuracy of the TNN model from Ref.\cite{Stoudenmire2016} has also been added to the plot with violet dashed-dot line.} One should note that the optimization landscapes and the gradient descent optimizers used for classical NNs and TNNs differ significantly. The DMRG training algorithm searches for the global minimum in the parameter space of multiple tensors in the \texttt{TNLayer}, whereas the \texttt{ADAM} optimizer, used for the NN model, searches within the parameter spaces of a single weight matrix in the \texttt{Dense} layer of the corresponding NN. For this reason, and to evaluate the accuracy of both models after utilizing all the training data only once, we chose a single epoch as the baseline for our experiment. However, the accuracy can be enhanced by increasing the number of epochs. For example the green-solid line in Fig.~\ref{Fig:mnist-accuracy-scaling} depict the accuracy of the NN model after $60$ epochs, which reached $0.976$. 

Next, we benchmarked the accuracy of the TNN model on the MNIST dataset with respect to the virtual bond dimension $chi$ as shown in Fig.~\ref{Fig:mnist-chi-scaling}-(a) for a fixed $N_{\rm MPO}=10$ and one training epoch. It is notable that the accuracy improves with an increase in the bond dimension. One should note that by increasing the bond dimension $\chi$, the number of parameters in the MPO tensors of the \texttt{TNLayer} also increases. Therefore, one can alternatively present the accuracy as a function of the parameters (see Fig.~\ref{Fig:mnist-chi-scaling}-(b)). This behavior is expected, as increasing the number of trainable parameters can positively impact accuracy, albeit at the cost of higher computational expense.

\begin{figure}
\centerline{\includegraphics[width=12cm]{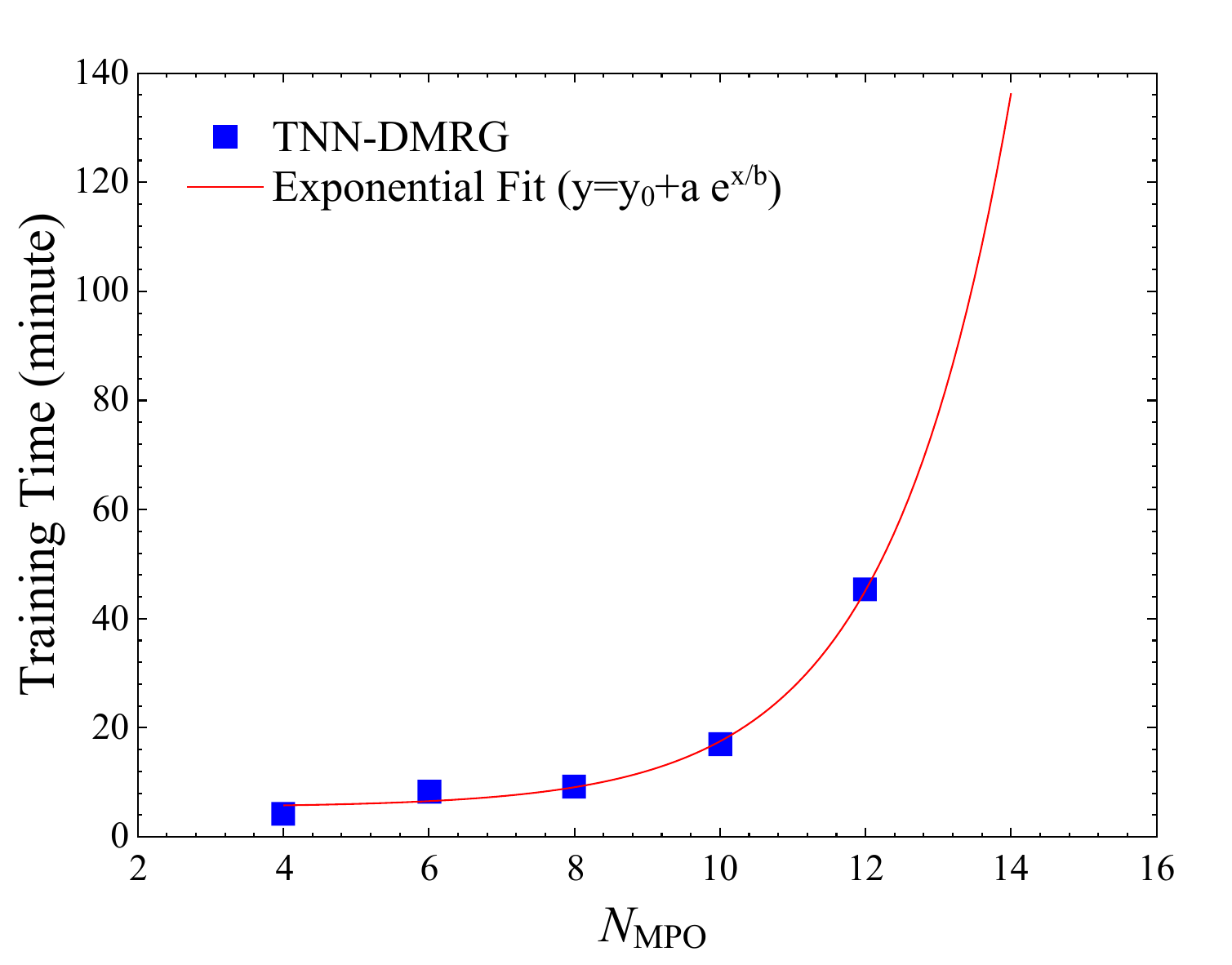}}
\caption{[Color online] Scaling of the MNIST classification training time for one epoch for with respect to the number of tensors, $N_{\rm MPO}$, in the \texttt{TNLayer} of the TNN model for one training epoch. The red line is the exponential fit to the training times with $y_0=5.42$, $a=0.03$ and $b=1.67$. The training time of the corresponding classical NN is $10$ seconds.}
\label{Fig:mnist-time-scaling}
\end{figure}

Finally, we have benchmarked the training time of the TNN model versus the classic NN model over one training epoch. Fig.~\ref{Fig:mnist-time-scaling} depicts the scaling of the training time for $60,000$ records in the MNIST dataset as a function of $N_{\rm MPO}$ for one sweep. While the training time for the classic NN model, using the \texttt{Adam} optimizer, is approximately $10$ seconds, the training time for the TNN model, which employs the sweeping DMRG-like algorithm, is on the order of minutes and increases exponentially with the number of tensors, $N_{\rm MPO}$. The red solid line in the figure represents an exponential fit to the TNN training times, indicating a sharp increase as the number of tensors grows.

{\color{blue} We conclude this section by noting that the sweeping DMRG-like training algorithm presented here served as a proof of concept and has not been optimized for performance during our benchmarks. Therefore, the TNN-DMRG, in its current CPU implementation, cannot surpass conventional gradient-based optimizers such as Adam. However, there is significant room for improvement, including efficient implementations of the tensor contractions on GPU or redesigning the building blocks of the optimizer. Additionally, the TNN models can be further optimized using any of the built-in optimizers in \texttt{TensorFlow} (see Ref.\cite{Patel2022} for an example).}

\section*{Conclusions and outlook}
\label{sec:conclude}

In this paper, we introduced a fully tensorized neural network model for deep learning. By replacing the trainable weight matrices of the fully connected dense layers of classic NNs with matrix product operators, we obtained tensor neural networks that are capable of modelling different machine learning tasks ranging from classification to regression. We further introduced a new entanglement-aware training algorithm based on DMRG and local gradient-descent updates for training the TNN models which act on a reduced parameter subspace obtained from the tensorization of trainable weights. Our TNNs are generic-purpose, i.e., they can be used for automatizing different ML models such as regression and classification. Our implementation further allows to construct hybrid architectures with a mixture of \texttt{Dense} and \texttt{TNLayer}s to build real instances of deep learning models.    

In order to show the performance and accuracy of the TNNs and DMRG-like training algorithm, we considered several deep learning models for linear and non-linear regression,  classification of labeled data as well as image recognition with MNIST handwritten digits. Our findings suggest that TNNs and DMRG-like training algorithms can serve as potential alternatives to conventional NN algorithms, applicable across various machine learning tasks. Most importantly, the DMRG-like training algorithm provides direct access to the singular values along the virtual dimensions of the trainable MPOs of \texttt{TNLayer}s, from which a measure of entanglement (correlation) between the features and model parameters can be computed.

Our TNN and DMRG-like algorithm suggest that tensor networks are closely related to neural networks. In fact, our approach opens the door for designing new numerical techniques for obtaining neural network representations of a quantum state and is a valuable tool to study the expressive power of quantum neural states. Furthermore, the ideas developed here can further be extended to other deep learning architectures such as convolutional neural networks \cite{martin-ramiro_boosting_2024, singh_tensor_2024} and their training algorithm.

Last but not least, tensor network compression of trainable weights can have potential applications in providing compact representations of large models, such as large language models (LLMs). We have recently demonstrated how TN compression can be effectively used for efficient parameter reduction in LLMs, such as Llama-7B \cite{tomut_compactifai_2024}, resulting in a significantly more compact model while maintaining a high level of accuracy and efficiency.

\section*{Acknowledgements}

S.S.J. acknowledges the support from Institute for Advanced Studies in Basic Sciences (IASBS), Donostia International Physics Center (DIPC) and Multiverse Computing.

\section*{Author contributions statement}

All authors contributed equally to the manuscript.

\section*{Data Availability} 

The datasets used and/or analyzed during the current study available from the corresponding author on reasonable request.

\bibliography{references}

\end{document}